\documentclass[]{aa}
\usepackage[utf8]{inputenc}

\usepackage{graphicx}
\graphicspath{{./}}
\usepackage{amssymb,amsmath,amsfonts}
\usepackage{txfonts}

\usepackage{epsfig}
\usepackage{color}
\usepackage[dvipsnames]{xcolor}
\usepackage{url}
\usepackage[]{hyperref}
\hypersetup{colorlinks=true,linkcolor=blue,citecolor=blue,filecolor=blue,
urlcolor=blue}

\usepackage{layouts}

\usepackage{natbib}
\bibpunct{(}{)}{;}{a}{}{,}

\begin{document}

   \title{The LOFAR Two Metre Sky Survey: Deep Fields}
   \subtitle{II. The ELAIS-N1 LOFAR deep field\thanks{
The data associated with this article are released at: 
\mbox{https://lofar-surveys.org}}}
   \titlerunning{The ELAIS-N1 LOFAR deep field}
   \author{
J.~Sabater\inst{1,2}\thanks{E-mail: jsm@roe.ac.uk}
\and
P.~N.~Best\inst{1}
\and 
C.~Tasse\inst{3,}\inst{4}
\and 
M.~J.~Hardcastle\inst{5}
\and 
T.~W.~Shimwell\inst{6}
\and 
D.~Nisbet\inst{1}
\and
V.~Jelic\inst{7}
\and
J.~R.~Callingham\inst{6,}\inst{8}
\and 
H.~J.~A.~R\"ottgering\inst{8}
\and
M.~Bonato\inst{9,}\inst{10,}\inst{11}
\and
M.~Bondi\inst{9}
\and 
B.~Ciardi\inst{12}
\and
R.~K.~Cochrane\inst{13}
\and
M.~J.~Jarvis\inst{14,}\inst{15}
\and
R.~Kondapally\inst{1}
\and 
L.~V.~E.~Koopmans\inst{16}
\and
S.~P.~O'Sullivan\inst{17}
\and
I.~Prandoni\inst{9}
\and
D.~J.~Schwarz\inst{18}
\and 
D.~J.~B.~Smith\inst{5}
\and
L.~Wang\inst{19,}\inst{16}
\and
W.~L.~Williams\inst{8}
\and
S.~Zaroubi\inst{16}
}

    \institute{
SUPA, Institute for Astronomy, Royal Observatory, Blackford Hill, 
Edinburgh, EH9 3HJ, UK
\and
UK Astronomy Technology Centre, Royal Observatory, Blackford Hill, 
Edinburgh, EH9 3HJ, UK
\and
GEPI, Observatoire de Paris, CNRS, Universite Paris Diderot, 5 place 
Jules 
Janssen, 92190 Meudon, France
\and
Department of Physics \& Electronics, Rhodes University, PO Box 94, 
Grahamstown, 6140, South Africa
\and
Centre for Astrophysics Research, School of Physics, Astronomy and 
Mathematics, University of Hertfordshire, College Lane, Hatfield AL10 9AB, UK
\and
ASTRON, the Netherlands Institute for Radio Astronomy, Postbus 2, 7990 
AA, Dwingeloo, The Netherlands
\and
Ru{\dj}er Bo\v{s}kovi\'c Institute, Bijeni\v{c}ka cesta 54, 10 000 Zagreb, 
Croatia
\and
Leiden Observatory, Leiden University, PO Box 9513, NL-2300 RA Leiden, 
The Netherlands
\and
INAF -- Istituto di Radioastronomia, via Gobetti 101, 40129 Bologna, Italy
\and
Italian ALMA Regional Centre, Via Gobetti 101, I-40129, Bologna, Italy
\and
INAF -- Osservatorio Astronomico di Padova, Vicolo dell'Osservatorio 5, 
I-35122, Padova, Italy
\and
Max Planck Institute for Astrophysics, Karl-Schwarzschild-Str. 1 D-85748 
Garching
\and
Harvard-Smithsonian Center for Astrophysics, 60 Garden St, Cambridge, MA 02138, 
USA
\and
Astrophysics, Department of Physics, Keble Road, Oxford, OX1 3RH, UK
Department of Physics \& Astronomy
\and
University of the Western Cape, Private Bag 
X17, Bellville, Cape Town, 7535, South Africa
\and
Kapteyn Astronomical Institute, University of Groningen, Postbus 800, 9700 AV 
Groningen, the Netherlands
\and
School of Physical Sciences and Centre for Astrophysics \& Relativity, Dublin 
City University, Glasnevin, D09 W6Y4, Ireland
\and
Fakult\"at f\"ur Physik, Universit\"at Bielefeld, Postfach 100131, 33501 
Bielefeld, Germany
\and
SRON Netherlands Institute for Space Research, Landleven 12, 9747 AD, Groningen, 
The Netherlands
}

   \date{Accepted ---; received ---; in original form \today}
   
   \abstract
{
The LOFAR Two-metre Sky Survey (LoTSS) will cover the full northern sky and, 
additionally, aims to observe the LoTSS deep fields to a noise level of 
\mbox{$\lesssim 10 \mu$Jy beam$^{-1}$} over several tens of square degrees in 
areas that have the most extensive ancillary data. This paper presents the 
ELAIS-N1 deep field, the deepest of the LoTSS deep fields to date. With an 
effective observing time of 163.7 hours, it reaches a root mean square (RMS) 
noise level of \mbox{$\lesssim$20 $\mu$Jy beam$^{-1}$} in the central region 
(and below \mbox{30 $\mu$Jy beam$^{-1}$} over 10 square degrees). The resolution 
is $\sim$6~arcsecs and 84862 radio sources were detected in the full area (68 
square degrees) with 74127 sources in the highest quality area at less than 3 
degrees from the pointing centre. The observation reaches a sky density of more 
than 5000 sources per square degree in the central region ($\sim$5 square 
degrees). We present the calibration procedure, which addresses the special 
configuration of some observations and the extended bandwidth covered (115 to 
177~MHz; central frequency 146.2~MHz) compared to standard LoTSS. We also 
describe the methods used to calibrate the flux density scale using 
cross-matching with sources detected by other radio surveys in the literature. 
We find the flux density uncertainty related to the flux density scale to be 
$\sim$6.5 per cent. By studying the variations of the flux density measurements 
between different epochs, we show that relative flux density calibration is 
reliable out to about a 3 degree radius, but that additional flux density 
uncertainty is present for all sources at about the 3 per cent level; this is 
likely to be associated with residual calibration errors, and is shown to be 
more significant in datasets with poorer ionosphere conditions. We also provide 
intra-band spectral indices, which can be useful to detect sources with unusual 
spectral properties. The final uncertainty in the flux densities is estimated to 
be $\sim$10 per cent for ELAIS-N1.
}

               
\maketitle


\defcitealias{Tasse2020}{Paper~I}
\defcitealias{Kondapally2020}{Paper~III}
\defcitealias{Duncan2020}{Paper~IV}

\section{Introduction}
\label{sec:intro}

Deep, wide-area radio surveys, especially when combined with high-quality 
pan-chromatic data from ultraviolet to far-infrared wavelengths, provide an 
unparalleled view of the evolving Universe. As radio emission is unaffected by 
dust absorption, the radio waveband offers an unbiased view of star-forming 
galaxies, which at low radio frequencies primarily emit due to non-thermal 
synchrotron emission associated with supernovae \citep[e.g.][]{Condon1992}. 
Radio observations also provide a unique insight into the growth of the 
supermassive black holes that can be found in the centres of massive galaxies. 
As well as providing a largely dust- and orientation-independent view of 
powerful active galactic nuclei (AGN), radio observations are the only way to 
reliably identify the low-luminosity `jet-mode' AGN, hosted by massive quiescent 
galaxies, the feedback from which is understood to play a critical role in 
regulating the growth of massive galaxies 
\citep{Best2005b,Bower2006,Croton2006,Best2012,Heckman2014}. 

Radio surveys have been used to study the mechanisms triggering AGN jets and 
their feedback \citep[e.g.][]{Best2012}, but deep observations are required to 
study their evolution through the history of the Universe \citep{Best2014}. The 
population of star-forming galaxies starts to dominate at lower radio 
luminosities and deep surveys are crucial to further understand their properties 
\citep{Gurkan2019}. The low-luminosity end of the far infrared to radio 
correlation can also be probed this way \citep{CalistroRivera2017}. However, the 
deep radio observations required to study these source populations at high 
redshifts are usually limited to pencil beam surveys 
\citep{Ciliegi2005,Bondi2007,Morrison2010, Murphy2017a,Owen2018}, and the 
deepest degree-scale surveys currently available only cover 2 square degrees 
\citep[e.g.][]{Schinnerer2007,Smolcic2017}. Although MIGHTEE \citep{Jarvis2016} 
will extend this at GHz frequencies.

The Low Frequency Array \citep[LOFAR;][]{vanHaarlem2013} combines a wide field 
of view with a high sensitivity and high angular resolution; this combination of 
capabilities enables deep, wide-area, high-fidelity radio surveys. The LOFAR Two 
Metre Sky Survey \citep[LoTSS;][]{Shimwell2017, Shimwell2019} is generating a 
wide-area survey of the sky covering the Northern Hemisphere. It reaches a 
sensitivity of \mbox{$\lesssim 100\,\mu$Jy beam$^{-1}$} with a full width at 
half maximum (FWHM) synthesized beam of $\sim$6~arcsecs and its first data 
release (DR1) covering 424 square degrees is already public. This first data 
release has already enabled the study of the low-frequency sky in unprecedented 
ways \citep{Croston2019,Gurkan2019,Hardcastle2019,Mahatma2019,Mingo2019, 
Mooney2019,Morabito2019,Sabater2019,Stacey2019,Wang2019}. The LoTSS DR1 has 
already probed the relation between the triggering of AGN and the stellar mass 
\citep{Sabater2019} showing that all the most massive galaxies present 
AGN-related radio emission once the observations are deep enough to detect it.

We are complementing the wide-area LoTSS survey with a series of deeper 
pointings, known as the LoTSS deep fields. The LoTSS deep fields ultimately aim 
to reach an RMS depth of \mbox{$\sim$10 $\mu$Jy beam$^{-1}$}, which is 
comparable to the deepest existing pencil-beam surveys, but will achieve this 
over a sky area of 30-50 square degrees. This is sufficient sky area to probe 
all cosmic environments at $z \gtrsim 1$, as well as to build up statistically 
meaningful samples of the rarer objects, such as low-luminosity AGN and 
starburst galaxies at high redshifts. It will have the sensitivity to detect 
Milky Way like galaxies at $z \gtrsim 1$ or strong starburst galaxies up to $z 
\sim 5$. The fields selected for the deep survey are those with the 
highest-quality degree-scale multi-wavelength data at declinations above 30 
degrees, where LOFAR has its highest sensitivity: Bo{\"o}tes 
\citep{Jannuzi1999}, the Lockman Hole \citep{Lockman1986}, and the European 
Large-Area ISO Survey-North 1 \citep[ELAIS-N1][]{Oliver2000} which is the 
subject of this paper. The coverage of wide areas in different lines of sight 
will help to minimize the effects of cosmic variance. The LoTSS deep fields will 
remain competitive even after the advent of the first phase of the Square 
Kilometre Array \citep[SKA;][]{Dewdney2009, Ekers2012}, as their sensitivity 
will reach below the expected confusion noise of SKA-low at these frequencies 
\citep{Prandoni2015,Zwart2015}. Finally, given the multi-epoch observations 
required by the LoTSS deep surveys, the data can be used to detect transients, 
source variability and stellar emission.

The calibration of the deep fields data presents several challenges. The errors 
in the primary beam models of a phased array like LOFAR and the 
direction-dependent ionospheric effects are addressed by the new calibration 
techniques developed by \citet{Tasse2020} \citepalias[hereafter][]{Tasse2020}. 
These third generation radio-interferometry calibration techniques allow us to 
reach depths very close to the thermal noise limit. \citetalias{Tasse2020} also 
presents the calibration of Bo{\"o}tes and the Lockman Hole fields. The 
management of the big data volumes and computing requirements were explored and 
solved using new computing infrastructures like the cloud \citep{Sabater2017} or 
advanced High Throughput Processing infrastructures and techniques 
\citep[e.g.][]{Mechev2017}. This first LoTSS deep fields data release already 
covers a sky area of $\gtrsim 30$~deg$^2$ (in Bo{\"o}tes, the Lockman Hole, and 
ELAIS-N1) to RMS depths \mbox{$\sim$30 $\mu$Jy beam$^{-1}$}. The astrometric 
precision achieved in the calibration is good enough to allow the cross-match 
with multi-wavelength counterparts to radio sources. This data enrichment has 
proven to be fundamental for the generation of value added science in LoTSS DR1 
\citep{Williams2019, Duncan2019}. \citet{Kondapally2020} 
\citepalias[hereafter][]{Kondapally2020} has performed a careful compilation of 
the multiwavelength data available for radio source cross-identification in the 
deep fields. Host galaxy identifications were found for over 97 per cent of the 
radio sources detected in the region of excellent multiwavelength photometry. In 
the future, WEAVE-LOFAR \citep{Smith2016} will obtain deep spectroscopic 
measurements for essentially all sources detected in the LoTSS deep fields but 
high quality photometric redshifts have been produced by \citet{Duncan2020} 
\citepalias[hereafter][]{Duncan2020}. That study provides photometric redshifts 
and stellar mass estimates for millions of sources in the fields, including 
faint radio sources and optical sources not detected in radio.

In this paper, which is the second in the LoTSS deep fields series, we present 
the ELAIS-N1 deep field. In Sect.~\ref{sec:data} we go through a description of 
the LOFAR observations and the additional radio data used in this study. 
Sect.~\ref{sec:calibration} describes the specific calibration process followed 
for ELAIS-N1, which differed in some respects from the standard LoTSS approach 
due to the adoption of a different observing set-up. This section also contains 
a discussion of the techniques used to carefully calibrate the flux density 
scale. We describe the final data products and catalogues in 
Sect.~\ref{sec:final_catalogue} while in Sect.~\ref{sec:discussion} a detailed 
study of the noise levels, flux density variability, extended sources, and 
spectral index are presented. Finally, Sect.~\ref{sec:conclusions} summarizes 
the work and highlights the main conclusions.


\section{The data}
\label{sec:data}

The ELAIS-N1 field \citep{Oliver2000} was chosen for its ample multi-frequency 
coverage. To allow deep observations, it was originally selected: a) to be at 
high ecliptic latitudes to minimize the impact of zodiacal light; b) to have 
low far infrared intensity to minimize the impact of galactic cirrus and; c) to 
avoid bright infrared sources that can saturate infrared observations. The deep 
multi-wavelength (150~nm -- 500~$\mu$m) data are covered by 
\citetalias{Kondapally2020} and here we focus in the radio observations.

\subsection{LOFAR observations}
\label{sec:lofar_obs}

The ELAIS-N1 data presented in this paper were taken in the LOFAR observation 
cycles 0, 2, and 4 (proposals LC0\_019, LC2\_024, and, LC4\_008 respectively) 
from May 2013 to August 2015. The different observations and datasets are 
summarized in Table~\ref{tab:observations}. The data observed in early LOFAR 
cycles (0 and 2) were taken jointly with the LOFAR Epoch of Reionization Key 
Science Project team, as a potential field for EoR studies \citep{Jelic2014}. 
Hence, the observing configuration was different from the standard LoTSS 
configuration. The observations of the ELAIS-N1 field were taken with the LOFAR 
high band antennas with a frequency ranging from 114.9 to 177.4 MHz. The phase 
centre of the main field is located at 16:11:00 +55:00:00 (J2000). The 
configuration of the antennas was `HBA\_DUAL\_INNER'. In this configuration the 
HBA antennas of the LOFAR core stations are split into two independent stations 
and the HBA tiles of the LOFAR remote stations are cropped to a shape similar to 
that of the core stations. This provides a uniform general shape of the primary 
beam over the entirety of the LOFAR stations in the Netherlands.

\begin{table*}
\centering
\caption{List of the ELAIS-N1 observations. The columns are the following: i) ID 
- Internal ID code of the dataset, ii) LOFAR observation ID - Standard LOFAR ID, 
iii) Cycle - LOFAR observing cycle, iv) Date - Date and time at which the 
observation started, v) Obs. time - Total length of the observation in seconds, 
vi) Deep - Flag indicating if the dataset was directly used in the final deep 
image (1 -- yes; 0 -- no).}
\label{tab:observations}
\begin{tabular}{cccccc}
  \hline
 ID & LOFAR ID & Cycle & Date & Obs. time & Deep \\
\hline
\hline
 000 & L133271 & 0 & 2013-05-12 20:19:48 & 28800.0 & 0 \\
 003 & L138664 & 0 & 2013-05-20 19:48:21 & 28800.0 & 0 \\
 005 & L138658 & 0 & 2013-05-26 19:24:46 & 28800.0 & 0 \\
 009 & L229064 & 2 & 2014-05-19 19:49:19 & 28805.8 & 1 \\
 010 & L229312 & 2 & 2014-05-20 19:46:23 & 28805.8 & 1 \\
 011 & L229387 & 2 & 2014-05-22 19:30:00 & 28805.8 & 1 \\
 012 & L229673 & 2 & 2014-05-26 19:30:00 & 28805.8 & 1 \\
 013 & L230461 & 2 & 2014-06-02 19:30:00 & 28805.8 & 1 \\
 014 & L230779 & 2 & 2014-06-03 19:30:00 & 28805.8 & 1 \\
 015 & L231211 & 2 & 2014-06-05 19:30:00 & 28805.8 & 1 \\
 016 & L231505 & 2 & 2014-06-10 19:50:00 & 26406.0 & 1 \\
 017 & L231647 & 2 & 2014-06-12 19:50:00 & 25198.0 & 1 \\
 018 & L232981 & 2 & 2014-06-27 20:05:58 & 17998.6 & 1 \\
 019 & L233804 & 2 & 2014-07-06 19:59:00 & 18001.0 & 1 \\
 020 & L345624 & 4 & 2015-06-07 20:11:00 & 27606.3 & 1 \\
 021 & L346136 & 4 & 2015-06-14 18:31:32 & 27606.3 & 1 \\
 022 & L346154 & 4 & 2015-06-12 20:11:00 & 27606.3 & 1 \\
 023 & L346454 & 4 & 2015-06-17 20:11:15 & 27606.3 & 1 \\
 024 & L347030 & 4 & 2015-06-19 17:58:00 & 27606.3 & 1 \\
 025 & L347404 & 4 & 2015-06-24 20:11:00 & 27606.3 & 0 \\
 026 & L347494 & 4 & 2015-06-26 20:11:00 & 27606.3 & 1 \\
 027 & L347512 & 4 & 2015-06-29 20:11:00 & 27606.3 & 1 \\
 028 & L348512 & 4 & 2015-07-01 20:11:00 & 24001.3 & 1 \\
 029 & L351576 & 4 & 2015-07-18 19:11:00 & 27606.3 & 0 \\
 030 & L366792 & 4 & 2015-08-07 18:11:00 & 27606.3 & 1 \\
 031 & L369530 & 4 & 2015-08-22 16:11:00 & 27606.3 & 1 \\
 032 & L369548 & 4 & 2015-08-21 16:11:00 & 27606.3 & 1 \\
\hline
\end{tabular}
\end{table*}

The observations contain additional data for either 3 (Cycle 4) or 6 (cycles 0 
and 2) flanking fields simultaneously observed with the ELAIS-N1 central region 
but with a much lower effective bandwidth ($6 \times 19$ or $3 \times 38$ 
spectral sub-bands in comparison to 371 sub-bands in the main target field). 
These data were not used, except for the dataset 000 in which the bandwidth was 
evenly distributed between the 7 target fields ($7 \times 69$ sub-bands each). 
In this case, we used the data of the flanking field whose centre was close to 
the calibrator source 87GB~160333.2+573543 (at 16:04:34.5 +57:28:01.7 in J2000) 
to assist with calibrating the flux density scale, as explained later in 
Sect.~\ref{sec:flux_scale}. The frequency limits for this dataset were similar 
to that of the other datasets but its frequency coverage was sparse for each of 
the fields.

Of the 371 spectral sub-bands observed in the central field, 320 correspond to a 
62.5 MHz frequency range centred at 146 MHz which was used for the deep images 
presented in this paper. The rest of the sub-bands correspond to a higher 
frequency band ranging from 179.4 to 189.2 MHz which was not considered for this 
study. The reduced size of the primary beam at those high frequencies, combined 
with the significant modification of the calibration pipeline that would be 
required to take them into account, contributed to the decision to study them at 
a later stage. We limited the analysis to the core and remote LOFAR stations but 
the observations also include data from the international stations that will be 
used in the future for sub-arcsecond imaging in the central part of this field 
\citep[][Sweijen et al. in prep.]{Jackson2016,Morabito2016}.

All the observations were preceded by a 5 to 10-minute run on the calibrator 
3C295 and succeeded by another 5 to 10-minute run on 3C380. The latter was the 
one selected for the calibration process because the pre-existing model for this 
source seemed to produce the best results. The model was in the flux density 
scale of \citep{Scaife2012}. The calibration and analysis of the data are 
presented in Sect~\ref{sec:calibration}.

\subsection{Additional radio surveys of ELAIS-N1}
\label{sec:other_surveys}

Additional radio data for the ELAIS-N1 are available from several large-area 
radio surveys and catalogues such as: the 87GB catalogue \citep{Becker1991, 
Gregory1991} at 4.85~GHz; the NRAO Very Large Array (VLA) Sky Survey 
\citep[NVSS;][]{Condon1998} at 1.4~GHz; the Faint Images of the Radio Sky at 
Twenty-Centimeters survey \citep[FIRST;][]{Becker1995} at 1.4~GHz; the Texas 
Survey of Radio Sources \citep{Douglas1996} at 365~MHz; the Westerbork Northern 
Sky Survey \citep[WENSS;][]{Rengelink1997} at 325~MHz; the Sixth Cambridge 
Survey of Radio Sources \citep[6C;][]{Hales1990} at 151~MHz; the GMRT 150~MHz 
All-sky Radio Survey \citep[TGSS;][]{Intema2017}; the VLA Low-frequency Sky 
Survey \citep[VLSS;][]{Cohen2007} at 74~MHz; the VLSS Redux 
\citep[VLSSr;][]{Lane2014} at 74~MHz; and the Eighth Cambridge Survey of Radio 
Sources catalogue \citep[8C;][]{Hales1995} at 38~MHz.

\begin{figure}[htbp]
   \centering
\includegraphics[width=\linewidth]{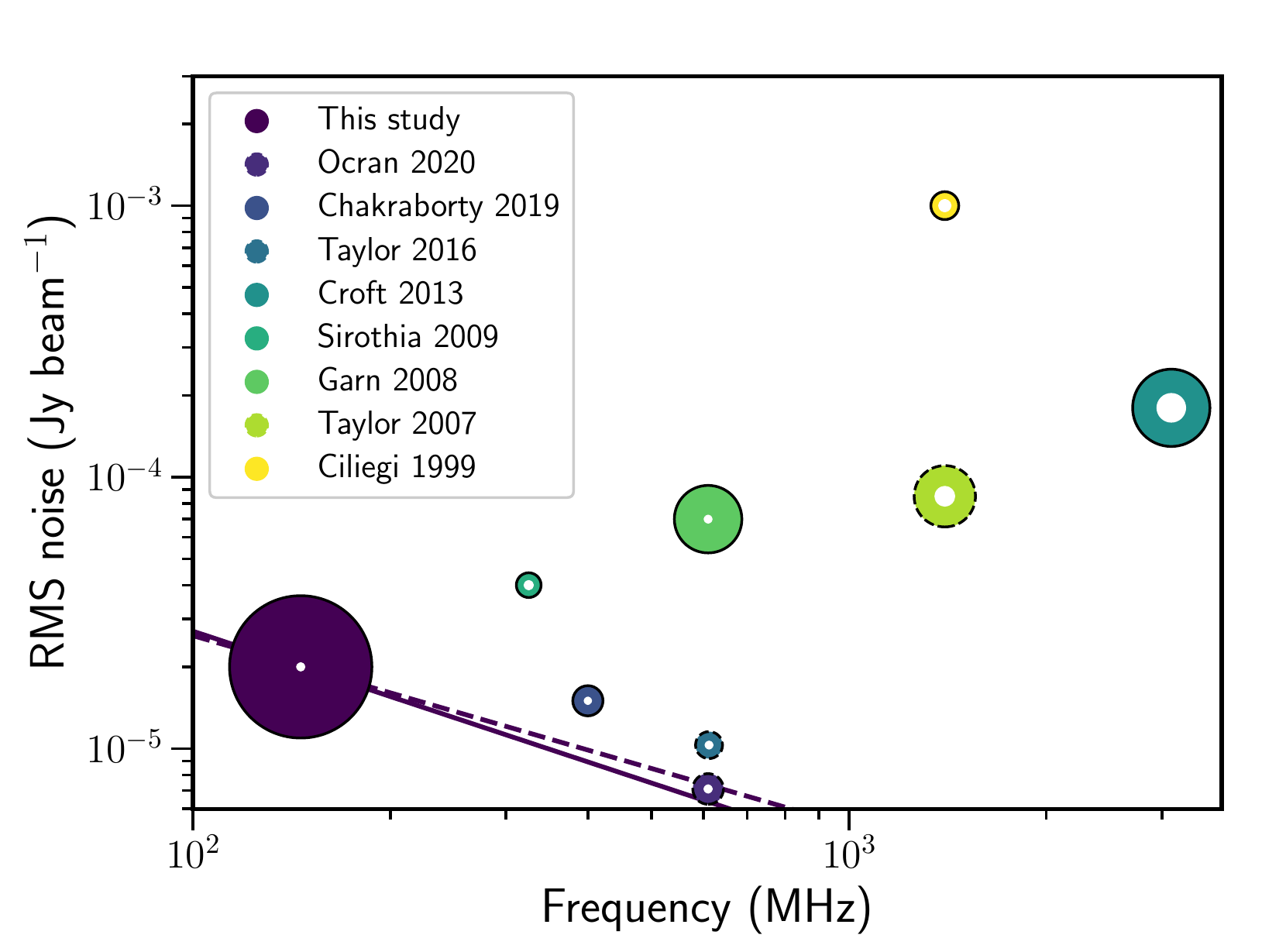}
   \caption{Depths and frequencies of radio surveys in the literature targeting 
the ELAIS-N1 field. The area of each circle is proportional to the area covered 
by the survey. The area of the inner dot is proportional to the resolution with 
smaller dots indicating better resolutions (ranging from 4.5 arcsecs up to 100 
arcsecs). The data were not found to be publicly available for the observations 
outlined with a dashed line \citep[those of][]{Taylor2007,Taylor2016,Ocran2020}. 
The solid line indicates a spectral index of 0.8 and the dashed line one of 
0.7.}
   \label{fig:comparison_surveys}
\end{figure}

Apart from these, it has also been observed to greater depths by targeted radio 
surveys. \citet{Ciliegi1999} observed the ELAIS-N1 field with the VLA at 1.4~GHz 
and detected 867 sources. Later, \citet{Taylor2007} observed the field in 
polarization at the same frequency but over a wider area, finding 786 compact 
sources. \citet{Garn2008} and \citet{Sirothia2009} observed the field with the 
Giant Metrewave Radio Telescope (GMRT) at 610 and 325 MHz respectively, 
detecting 2500 and 1286 sources. \citet{Croft2013} observed several fields 
including ELAIS-N1 with the Allen Telescope Array at 3.1 GHz detecting $\approx$ 
200 sources in the area. \citet{Taylor2016} studied the orientation of extended 
radio sources based on new data taken by the GMRT at 612 MHz. 
\citet{Chakraborty2019} observed ELAIS-N1 with the upgraded GMRT at frequencies 
between 300 and 500~MHz resulting in a catalogue of 2528 sources. Finally, the 
most recent observation of ELAIS-N1 was taken by \citet{Ocran2020} with the GMRT 
at 610~MHz and produced a catalogue of 4290 sources. In 
Fig.~\ref{fig:comparison_surveys} we show the parameters of these previous 
observations targeting ELAIS-N1 compared to our study, focusing mainly on the 
RMS noise achieved, the area covered, the resolution, and the public 
availability of the data. As shown in the figure, if we consider a canonical 
spectral index of $\alpha = 0.7$ \citep[for $S_\nu \propto 
\nu^{-\alpha}$;][]{Condon2002} our catalogue reaches greater depth than any of 
the previous catalogues as well as covering a substantially wider area than the 
other deeper surveys, at higher angular resolution than most.  

\section{Data calibration}
\label{sec:calibration}

The calibration and processing of the ELAIS-N1 data presents several challenges. 
The first one is the data size and computing power required. Each dataset can 
amount to 80 TB and its calibration requires several CPU-years which makes 
critical the use of a High Throughput Computing (HTC) or High Performance 
Computing (HPC) system for their processing \citep{Sabater2017}. The second one 
is the correction of the especially strong effect of the ionosphere on the 
signal at low frequencies \citep[e.g.][]{Intema2009}. This correction required 
the development of new calibration pipelines and software which are outlined in 
\citetalias{Tasse2020} of this series.

\label{sec:data_volume}
The volume of data is reduced as it is processed through the pipeline. The 
original amount of data corresponding to the observations listed in 
Table~\ref{tab:observations} is $\approx$2 PB. An individual Cycle 2 observation 
of 8 hours amounts to $\approx$80 TB as outputted from the observatory in raw 
format. This volume was reduced to about 12 TB by pre-processing (see 
Sect.~\ref{sec:pre-processing}) and to 4 TB after the first calibration step 
(\textsc{prefactor}; see Sect.~\ref{sec:DIcal}). The final calibrated 
measurement set \citep[MS;][]{vanDiepen2015} for each observation occupies 
$\approx$0.9 TB which can be further compressed to about 80 GB by averaging and 
removing redundant data and using \textsc{dysco} compression 
\citep{Offringa2016a}. Finally, the deep image of the ELAIS-N1 field that 
corresponds to 163.7 hours of observation amounts to a mere 1.5 GB. Hence, the 
data rates expressed as a fraction of total observing time follow this sequence 
as the calibration progresses: 2.7~GB/s $\rightarrow$ 350~MB/s $\rightarrow$ 
88~MB/s $\rightarrow$ 31~MB/s $\rightarrow$ 2.6~MB/s $\rightarrow$ 2.5~kB/s. 

\subsection{Pre-processing}
\label{sec:pre-processing}

The calibration of data taken in different cycles was performed slightly 
differently. Cycle 0 and 2 data were pre-processed by the Epoch of Reionization 
LOFAR Key Science Project team in order to remove radio-frequency interference 
(RFI) and demix the contribution of bright off-axis sources \citep{vTol2007}. 
Each of the 320 sub-bands into which the frequency range was divided consisted 
of 64 channels. To minimize bandpass effects, the two upper and two lower 
channels were removed and the remaining 60 channels averaged by a factor 4 to a 
final 15 channels per sub-band. The scan time was averaged by a factor 2 up to 2 
seconds. 

The Cycle 4 data were observed in a configuration similar to  that of the rest 
of LoTSS Survey data \citep{Shimwell2017, Shimwell2019} but maintaining the 
extended bandwidth. In this case the channels at the edge of the sub-bands were 
not discarded and the data were averaged to 16 channels per sub-band and 2 
seconds per scan.

\subsection{Direction-independent calibration}
\label{sec:DIcal}

The first step of the calibration for all the datasets used for the final deep 
image was made with the software \textsc{prefactor}\footnote{ Development: 
https://github.com/lofar-astron/prefactor; Documentation: 
https://www.astron.nl/citt/prefactor/} or earlier versions of this pipeline. The 
pipeline is described by \citet{deGasperin2019} and earlier versions are 
outlined in \citet{vanWeeren2016a} and \citet{Williams2016}. The observation of 
a calibrator source before or after the main target is used to obtain an 
estimation of several calibration parameters. In our case we used 3C380 as our 
calibrator source. The calibrator data are first flagged for RFI with 
\textsc{AOflagger} \citep{Offringa2012} and for problematic antennas or periods 
of time. The data are then averaged to 4 seconds of time resolution and 1 
channel per sub-band. Then the following effects are estimated and corrected 
for, in order: a) the polarization misalignment introduced by the station 
calibration tables; b) the Faraday rotation; c) the amplitude bandpass; d) the 
clock offset originating mainly from the drift of the clock in the remote 
stations and; e) the ionospheric direction-independent delays.

Once the calibration tables are generated, a first estimation of the effect of 
the ionosphere \citep{Mevius2016} is made with \textsc{RMextract}\footnote{ 
https://github.com/lofar-astron/RMextract; https://ascl.net/1806.024} 
\citep{RMextract}. The data are flagged for RFI and the parts of the data that 
are heavily affected by A-team sources\footnote{The `A-team' sources are 
extremely bright sky sources that can affect the observations even when they are 
far off-axis. They are usually designated as the name of the constellation where 
they are found and the suffix `A'. The main sources that are considered are 
Cygnus A, Cassiopeia A, Virgo A, and Taurus A but also Hercules A, Hydra A, and 
some bright 3C sources.} (data with an A-team predicted flux density 
contribution higher than 5~Jy) are flagged. After that, the calibration 
solutions obtained for the calibration target combined with the preliminary 
ionospheric effect estimations are applied to the target field. The data are 
then concatenated in chunks of 10 sub-bands which increases their 
signal-to-noise ratio (S/N) for the final phase gain calibration. A model of the 
field is required for this step. For ELAIS-N1, the sky model was built up 
historically from early Cycle observations. Appendix~\ref{sec:skymodel} details 
the creation of that model. The final outputs of the pipeline are the corrected 
$uv$ data, an estimated model of the sky and several diagnostics plots to check 
the accuracy of the calibration.

A primitive version of \textsc{prefactor} was used for Cycle 0 datasets which 
were used as test-beds for the development of the first direction-independent 
attempts. As a preparation for the final deep imaging calibration combining all 
the datasets, the \textsc{prefactor} calibration pipeline was applied to Cycle 2 
data in the Amazon Web Services (AWS) cloud premises \citep[see 
Appendix~\ref{sec:infrastructure} and ][]{Sabater2017}. Cycle 4 data were run 
through the \textsc{prefactor} pipeline set up in SURF-Sara \citep{Mechev2017}. 
One technical difficulty arose from the uneven spacing between channels in Cycle 
0 and 2 data after the sub-bands were combined. The frequencies for the 
different bands were edited to be spaced homogeneously which changed some of 
them by factors of a few per thousand. This modification permitted us to run the 
standard LOFAR software on the data without introducing significant biases into 
the calibration. All the pre-calibrated datasets were transferred to the compute 
cluster \textsc{Cuillin} at the University of Edinburgh for further processing. 

\subsection{Direction-dependent calibration}
\label{sec:DDcal}

The third generation calibration and imaging techniques for radio astronomy 
involve the estimation and compensation for direction-dependent effects 
\citepalias{Tasse2020}. The development of the solver 
\textsc{KillMS}\footnote{https://github.com/saopicc/killMS} \citep{Tasse2014a, 
Tasse2014b, Smirnov2015} and the imager 
\textsc{DDFacet}\footnote{https://github.com/saopicc/DDFacet} \citep{Tasse2018} 
constituted a big step forwards in the calibration effort. A pipeline that 
leverages these tools, named 
\textsc{DDF-pipeline}\footnote{https://github.com/mhardcastle/ddf-pipeline}, was 
developed. Versions 1 and 2 of the pipeline are explained in detail in Sect.~3 
and 4 of \citetalias{Tasse2020} respectively, and the reader is refered to that 
paper for full details; here a brief summary is provided.

\textsc{DDF-pipeline} works on the data that has been calibrated by 
\textsc{prefactor} in a direction-independent manner. A subset of the data, 
composed of the central 60 sub-bands, is imaged and the field is divided in 
facets\footnote{A facet is a sky polygon associated to a given direction (or 
coordinates) in the sky. The facet is associated to a set of solutions and 
parameters that are considered to be valid within this region of the sky. A 
tessellation of A big field is tessellated into smaller facets in order to 
consider direction-dependent effects.} using a clustering process. This subset 
is calibrated and imaged to produce a preliminary direction-dependent sky model 
and a deconvolution mask. Due to the difficulty to obtain a true flux density 
calibration from LOFAR data alone, these steps also include a bootstrap process 
to determine the flux density scale corrections \citep{Hardcastle2016}. In 
\textsc{DDF-pipeline v2}, the corrections are derived by comparison with matched 
WENSS and NVSS sources, using an empirical mean spectral index. With the new 
improved and flux-corrected model and the preliminary deconvolution masks, the 
full bandwidth data are subsequently processed. The final steps consist of a 
direction-dependent calibration followed by a direction-independent calibration, 
an imaging step and a further set of direction-dependent calibrations (slow and 
fast) designed to recover as much extended emission as possible. The final step 
is an imaging run including the solution and correction of the astrometric 
errors for each individual facet.

The run combining all the datasets was prepared by running the last version of 
the \textsc{DDF-pipeline} on dataset 015 which had a low noise level. The output 
model and mask of this run was used as the final input of the pipeline run on 
the combined set of data. In this case, the model and mask were used as an input 
to do a direction-independent and direction-dependent calibration of each 
dataset. After that they were imaged all together with the appropriate 
calibration solutions applied. During the imaging step the mask was updated to 
ensure that faint sources which were only detected in the combined dataset were 
deconvolved. Cycle 0 datasets and datasets 025 and 029 were excluded from the 
final image due to the poor noise levels or problems with the data. The final 
list of datasets used is flagged `1' in the column `Deep' of 
Table~\ref{tab:observations}.

\subsection{Pipeline data products}
\label{sec:data_products}

\begin{figure*}[!htbp]
   \centering
\includegraphics[width=0.85\linewidth]{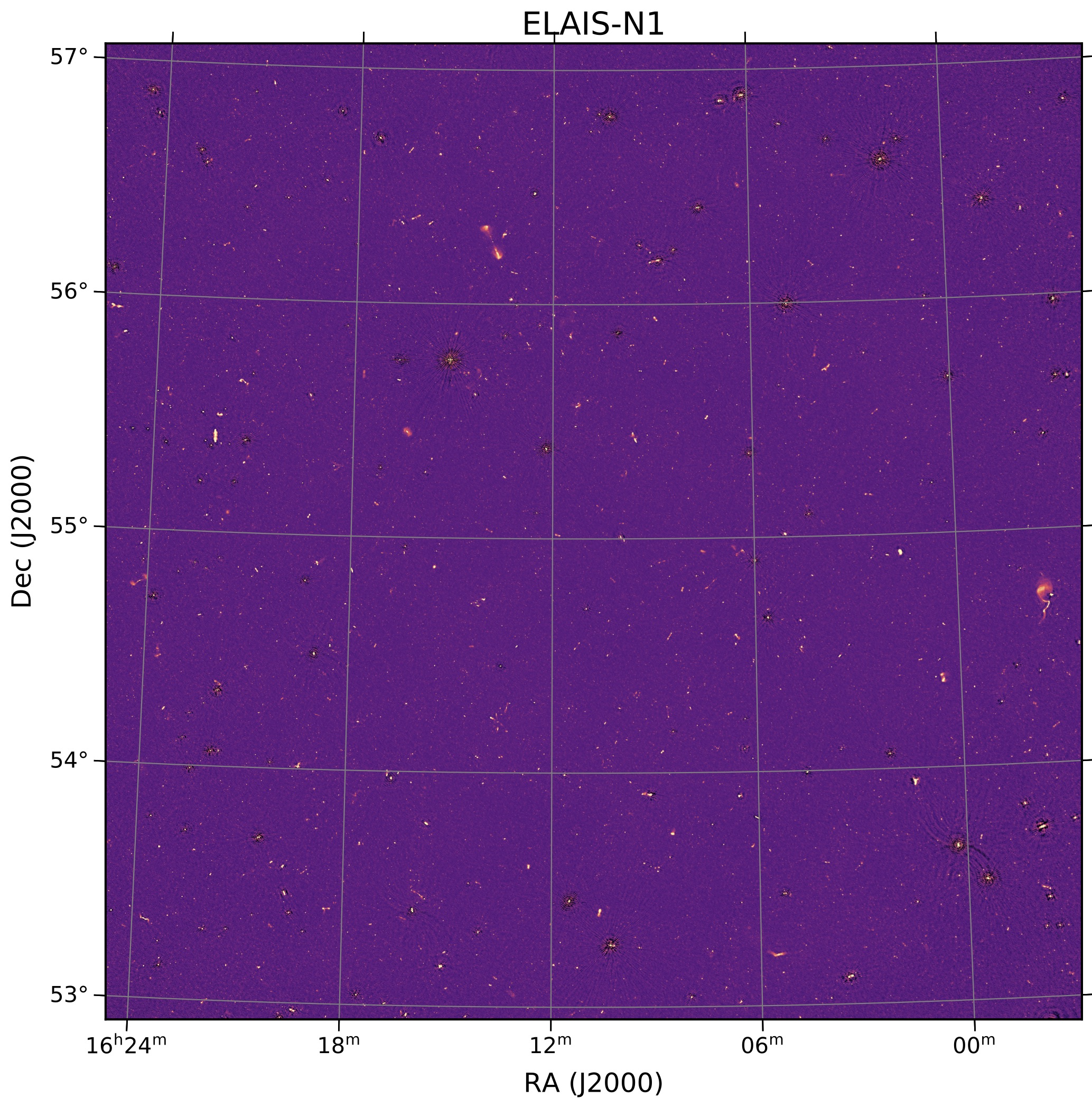}
\includegraphics[width=0.422\linewidth]{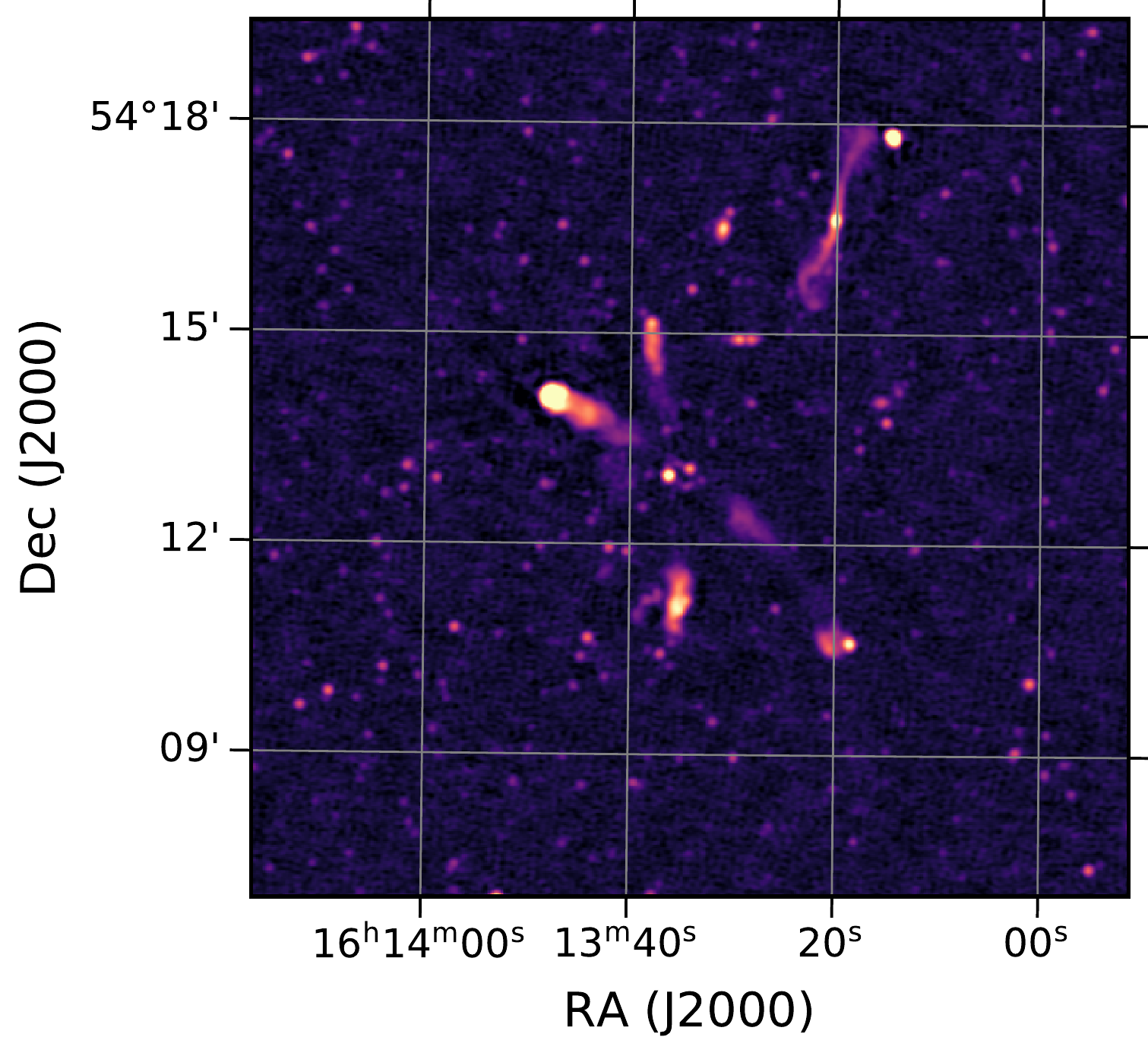}
\includegraphics[width=0.422\linewidth]{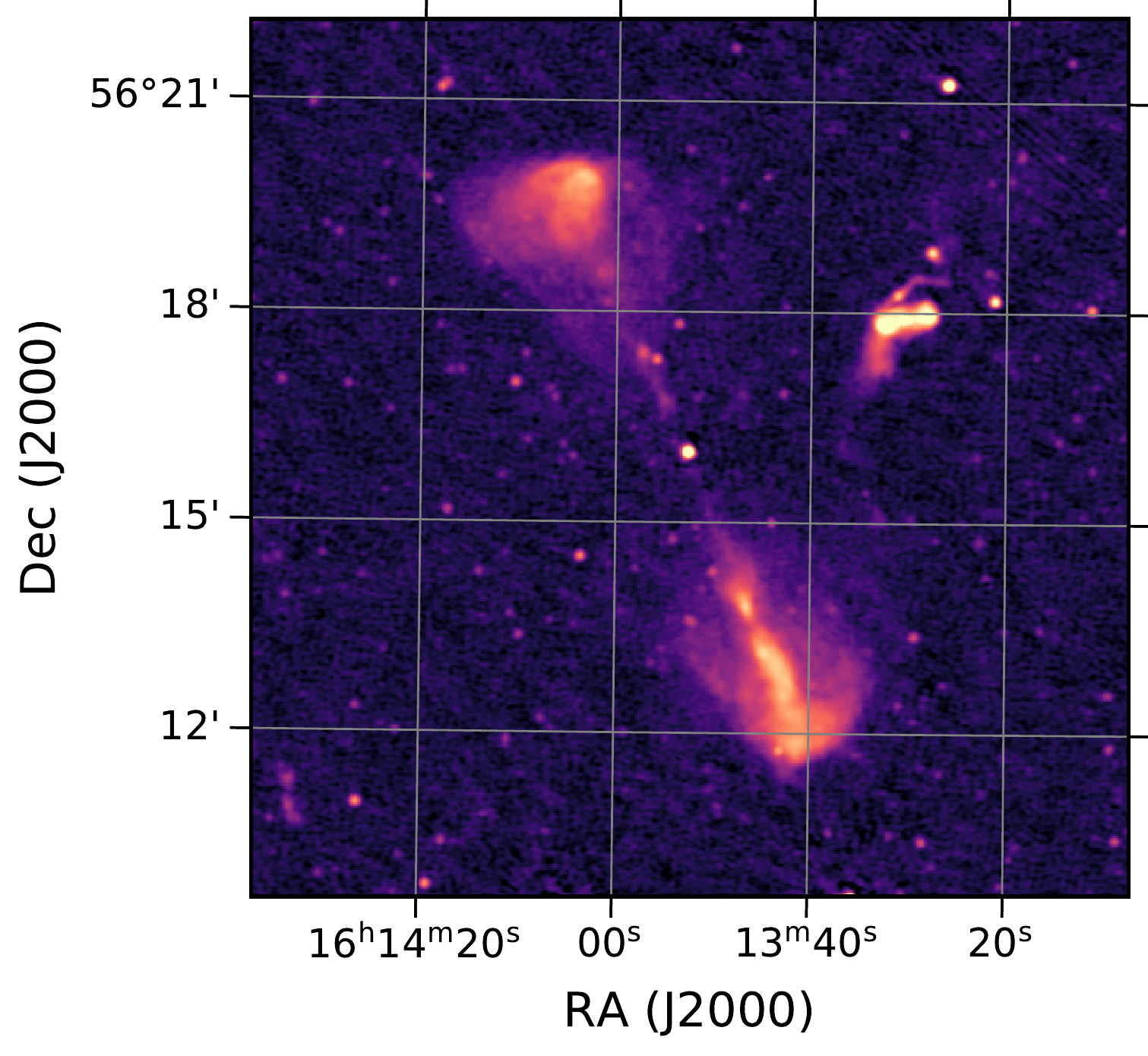}
   \caption{Final deep image of the central region of ELAIS-N1 field. The top 
panel shows a wide-field view containing about one third of the total area. The 
two lower panels show zoomed-in regions (each around 0.1 per cent of the full 
image area), showing the high image quality for extended sources and sensitivity 
to faint sources. All images are Stokes I, with 6 arcsecs resolution. The 
full-field image contains more than 85000 sources.}
   \label{fig:image}
\end{figure*}

The final run of the \textsc{DDF-pipeline} produced the following set of data 
products: a) Solutions corresponding to each facet and dataset composed of quick 
and slow smoothed solutions \citepalias[see Sect.~4.3 in ][]{Tasse2020}; b) deep 
high resolution (6 arcsecs) Stokes I image of the field which is shown in 
Figure~\ref{fig:image} and; c) deep low resolution (20 arcsecs) Stokes I image 
and Stokes V uncleaned image of the field. The data were also divided in 
frequency bands for further study of the consistency of the data and the 
intra-band spectral indices. Three bands with a set of frequencies 
nearly-equivalent to the ones used in the LoTSS wide area survey were produced, 
additionally three different bands covering the full extended spectral range of 
the ELAIS-N1 data were also produced. The frequency ranges used are shown in 
Table~\ref{tab:bands}. Bands 0, 1, and 2 correspond to the LoTSS wide-field 
survey configuration and bands X, Y, Z are extended. The spectral coverage of 
these extended bands was selected empirically in order to produce a similar 
median signal-to-noise level in the three bands for the sources in the central 2 
degree radius region of the ELAIS-N1 field.

\begin{table*}
\centering
\caption{Band frequency ranges of different datasets available in the data 
release. The table contains the following columns: i) The name assigned to the 
band configuration: 0--2 are designed to closely match the three LoTSS bands, 
while X--Z offer higher-sensitivity wider options; ii) Initial frequency of the 
band; iii) Final frequency of the band; iv) Central frequency of the band; v) 
Total bandwidth of the band; and vi) The range of blocks of 10 sub-bands 
combined to form the band.}
\label{tab:bands}
\begin{tabular}{cccccc}
\hline
 Band name & Initial freq. & Final freq. & Central freq. & Bandwidth & Combined 
bands \\
 & MHz & MHz & MHz & MHz &  \\
\hline
\hline
 0 & 120.80 & 136.42 & 128.61 & 15.62 & 3 to 10 \\
 1 & 136.42 & 152.05 & 144.23 & 15.62 & 11 to 18 \\
 2 & 152.05 & 167.67 & 159.86 & 15.62 & 19 to 26 \\
 X & 114.94 & 140.33 & 127.63 & 25.39 & 0 to 12 \\
 Y & 140.33 & 157.91 & 149.12 & 17.58 & 13 to 21 \\
 Z & 157.91 & 177.44 & 167.67 & 19.53 & 22 to 31 \\
Full & 114.94 & 177.44 & 146.19 & 62.50 & 0 to 31 \\
\hline
\end{tabular}
\end{table*}

The solutions were later used to produce Stokes I and V images corresponding to 
the individual datasets. Low resolution (20 arcsecs) and very low resolution 
(4.3 arcmin) Stokes QU datacubes of the individual datasets were also obtained 
for further processing \citep[see][]{HerreraRuiz2020}. Those datacubes are split 
in 800 or 640 frequency channels in the case of Cycle 2 or 4 datasets 
respectively due to their different frequency configuration.

Preliminary ELAIS-N1 catalogues were produced from the high resolution Stokes-I 
images of the deep field, the spectral bands, and the individual dataset images. 
The sources were extracted from the images using \textsc{PyBDSF} 
\citep{Mohan2015}. \textsc{PyBDSF} extracts a catalogue of sources that can be 
composed of either individual or multiple Gaussians. It takes into account 
different scale effects by using wavelets. An RMS noise distribution image is 
also produced by the software. The parameters used to extract the sources are 
detailed in Appendix~\ref{sec:pybdsfparameters}. More than 80000 sources were 
extracted from the deep image (see Sect.~\ref{sec:final_catalogue} for more 
details).

\subsection{Flux scale calibration}
\label{sec:flux_scale}

The preliminary version of the catalogues was compared to the initial sky model 
derived in Appendix~\ref{sec:skymodel}. This indicated a possible 
over-estimation of the flux density scale by \textsc{DDF-pipeline}. The high 
quality of existing multi-frequency radio data in the ELAIS-N1 field should 
permit a higher accuracy flux density calibration than the default 
\textsc{DDF-pipeline} methods which rely on cross matching to other very wide 
area radio surveys. We therefore used two methods to test the flux density scale 
of ELAIS-N1: a) a comparison of the flux density of a calibrator source within 
the field, and b) a cross-match with additional radio catalogues covering the 
area.

The calibrator 87GB~160333.2+573543 is observed within the main target field, 
and close enough to the pointing centre not to be significantly affected by 
primary beam correction problems (2.6 degrees). Accordingly, we compared its 
flux density with that estimated during the construction of the initial sky 
model in Appendix~\ref{sec:skymodel}. We measure a surface brightness of 
\mbox{5.58 Jy beam$^{-1}$} and the expected flux density according to the model 
is \mbox{4.44 Jy} as shown in Appendix~\ref{sec:skymodel}. As the source is not 
extended at 6 arcsec resolution this implies that the flux density is 
overestimated by 26 per cent and a correction factor of 0.796 should be applied. 
However, as this estimate is based on a single source whose properties as 
calibrator are not well determined we looked into more robust methods that used 
additional data.

\begin{figure*}[htbp]
  \centering
  \includegraphics[width=0.48\linewidth]{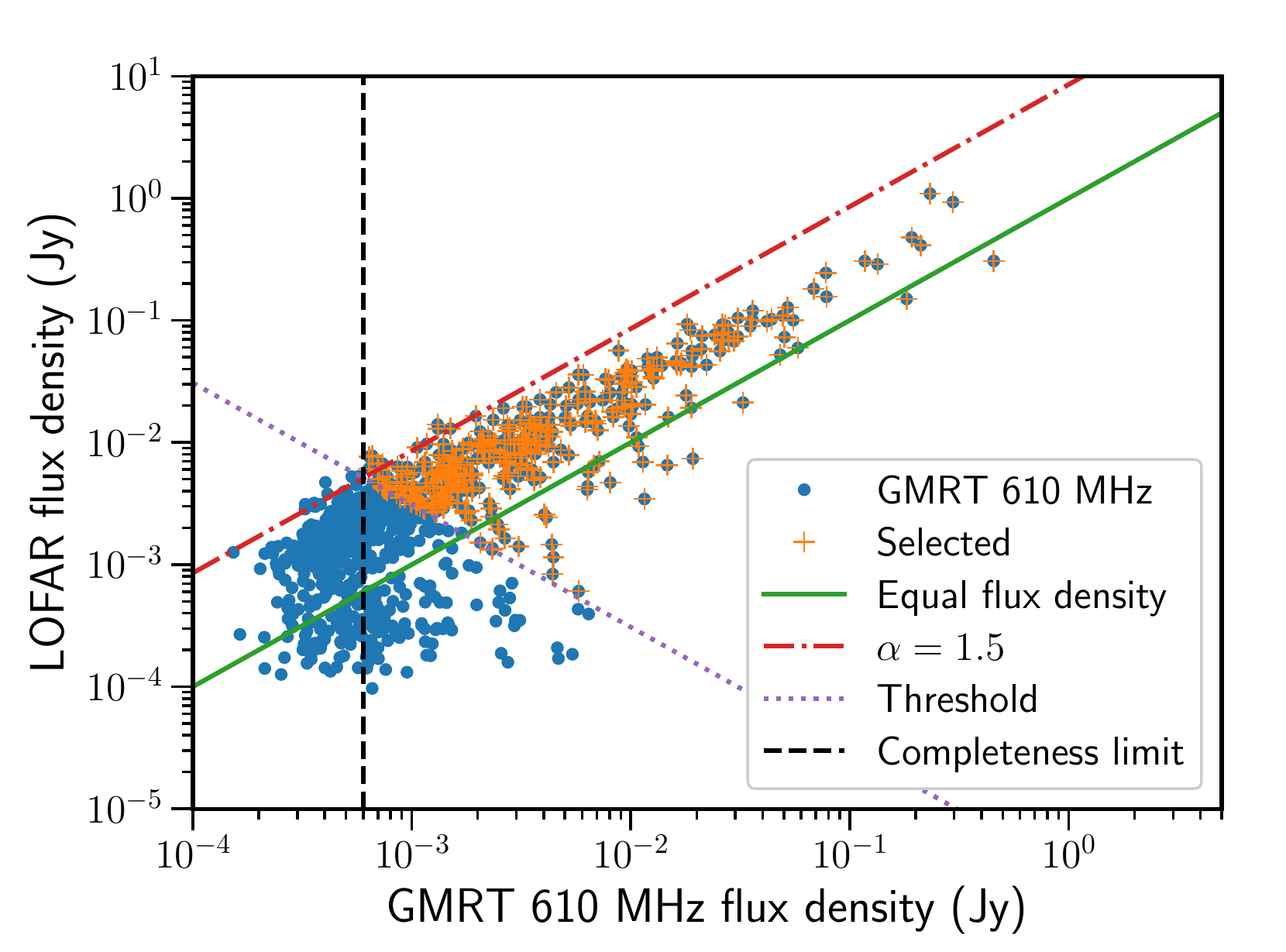}
  \hfill
  \includegraphics[width=0.48\linewidth]{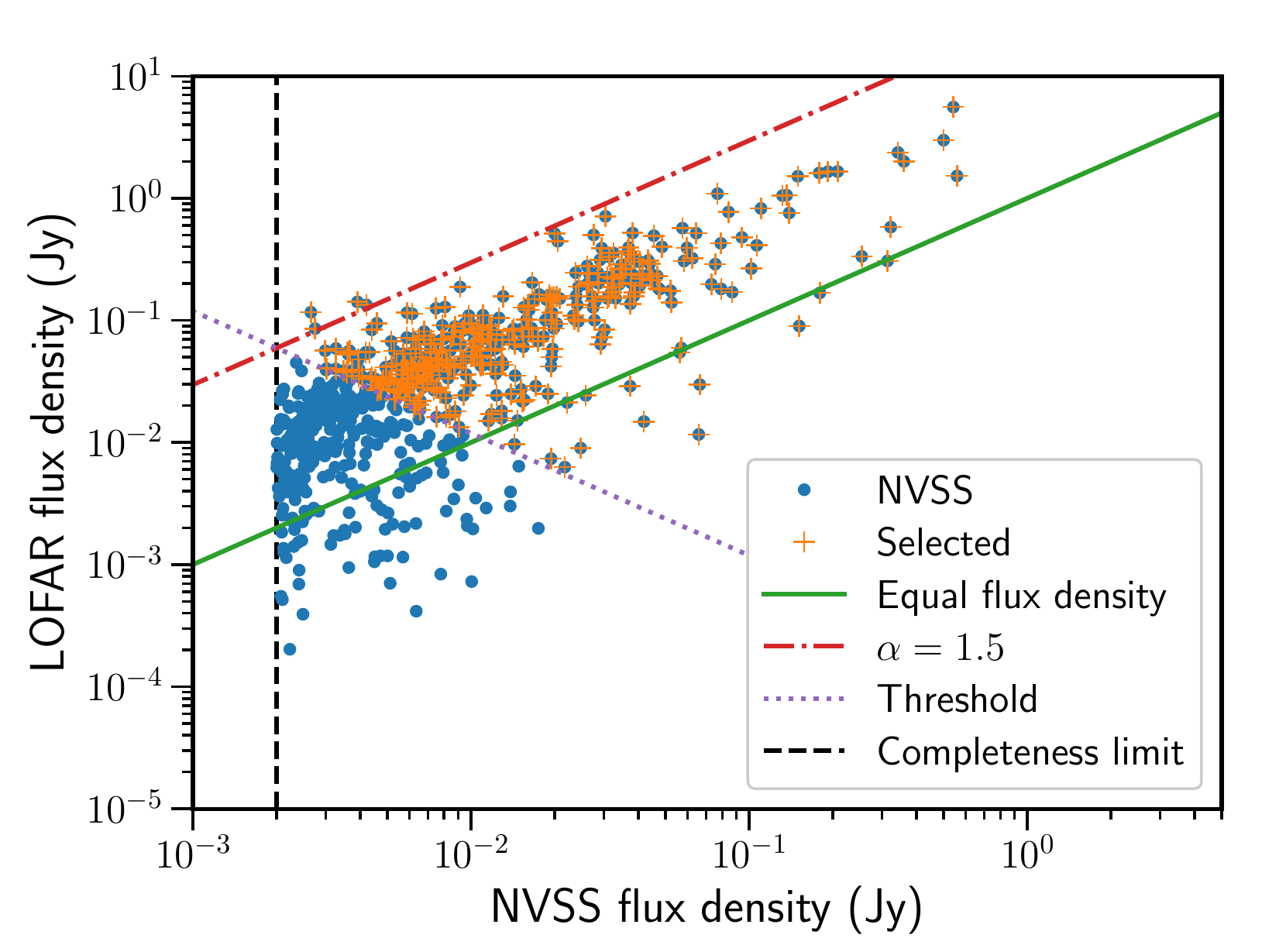}
  \caption{Selection of sources for the cross-match for the GMRT 610 MHz (left 
panel) and the NVSS (right panel) samples. All the cross-matched sources are 
shown as blue dots and the finally selected ones are marked with an orange 
cross. The completeness limit of the survey is shown as a vertical dashed line. 
The line at which the flux densities are equal is shown as reference as a solid 
green line. The locus of sources with a spectral index of 1.5 is shown as a 
dash-dotted red line. The point at which this line and the survey completeness 
line cross is used as a reference for the selection threshold in the product of 
flux densities (see text) which is marked as a purple dotted line.}
  \label{fig:flux_selection}
\end{figure*}

The flux density of sources in common between our catalogue and additional 
external radio catalogues was compared as well. We compared with the VLSSr at 
74~MHz, the TGSS at 150~MHz, the 6C at 151~MHz, the ELAIS-N1 GMRT survey at 
325~MHz, WENSS at 350~MHz, the ELAIS-N1 GMRT survey at 610~MHz, NVSS at 1.4~GHz, 
and FIRST at 1.4~GHz. The external catalogues were cross-matched with the LOFAR 
catalogue, within the inner 3 degrees radius from the LOFAR pointing centre. The 
catalogues were on (or, where necessary, adjusted to) \citet{Baars1977} flux 
density scale for the higher frequencies and for the lower ones on 
\citet{Scaife2012} flux density scale which is set to be compatible with 
\citet{Baars1977} but more accurate at frequencies below $\sim300$~MHz. We also 
applied some constraints to avoid the introduction of biases produced by the 
different effective depths and angular resolutions of the surveys. The exact 
numerical parameters used are shown in Table~\ref{tab:filter}. In order to 
ensure a fair comparison of surveys with different angular resolutions, with no 
contamination by neighbouring sources, only isolated LOFAR sources were 
considered. An offset limit to the nearest neighbour LOFAR source was 
empirically chosen. This offset depended on the angular resolution of the 
comparison survey with lower resolution surveys requiring larger values to avoid 
contamination from neighbours. Additionally, the maximum cross-match distance 
between the LOFAR and the survey source was empirically set depending on the 
resolution of the matched survey. Surveys with lower resolution require larger 
cross-match distances. To avoid resolution selection effects, only compact LOFAR 
sources were selected by restricting their maximum size. Higher resolution 
surveys require the consideration of less extended LOFAR sources (e.g. because 
FIRST does not have the same surface brightness sensitivity to extended 
structures that LOFAR has). Incompleteness effects were minimized by considering 
the completeness limit of each comparison survey \citep[e.g.][]{Nisbet2017}. A 
survey dependent minimum flux density threshold was applied. Setting a threshold 
in only one of the cross-matched surveys may introduce a bias towards sources 
with high absolute values of their spectral indices. Therefore, an additional 
constraint was introduced as a threshold in the product of flux densities of our 
LOFAR measurement and each comparison survey. The threshold was set to be the 
flux density that a source at the completeness limit of the comparison survey 
multiplied by the LOFAR flux density that such a source would have for a 
spectral index of 1.5 (except for the lower frequency survey VLSSr, where a flat 
spectral index was used). The two panels of Fig.~\ref{fig:flux_selection} show 
this threshold as well as the completeness threshold for two of the surveys 
used. Most of the optimal parameters shown in Table~\ref{tab:filter} were 
empirically determined by \citet{Nisbet2017}. All these parameters were used to 
filter the LOFAR sample down to that which would produce an unbiased comparison 
for each comparison survey.

\begin{table*}
\centering
\caption{Cross-match parameters and figures. Parameters used for the cross-match 
with external surveys and their filtering. The columns are the following: i) 
Survey name, ii) Maximum cross-match distance for the survey, ii) Survey 
completeness limit, iii) Maximum size of the major axis used for filtering, iv) 
Minimum distance to the nearest neighbour LOFAR source used for filtering the 
isolated sources, v) Value of the selection threshold in the product of the flux 
densities (see text), vi) Number of sources selected after the cross-match and 
filtering.}
\label{tab:filter}
\begin{tabular}{lcccccc}
\hline
 Survey & 
 Max. cross-match & 
 Survey Flux  & 
 Max. major&
 Min. distance to & 
 Flux product & 
 N \\
 & 
 radius & 
 dens. limit & 
 axis size &
 nearest neighbour & 
 threshold & 
  \\
   & 
 (arcsecs) & 
 (mJy) & 
 (arcsecs) & 
 (arcsecs) & 
 (mJy) & 
  \\
\hline
\hline
 VLSSr & 10 & 530 & 25 & 60 & 318 & 13 \\
 TGSS & 5 & 65 & 20 & 60 & 64 & 74 \\
 6C & 15 & 100 & 20 & 60 & 49 & 12 \\
 GMRT 325 MHz & 8 & 1.2 & 20 & 40 & 0.8 & 115 \\
 WENSS & 6 & 5.5 & 10 & 10 & 29 & 139 \\
 GMRT 610 MHz & 10 & 0.6 & 10 & 10 & 1.8 & 338 \\
 NVSS & 10 & 2 & 15 & 20 & 11 & 338 \\
 FIRST & 4 & 2 & 10 & 10 & 11 & 310 \\
\hline
\end{tabular}
\end{table*}

\begin{figure*}[htbp]
   \centering
\includegraphics[width=\linewidth]{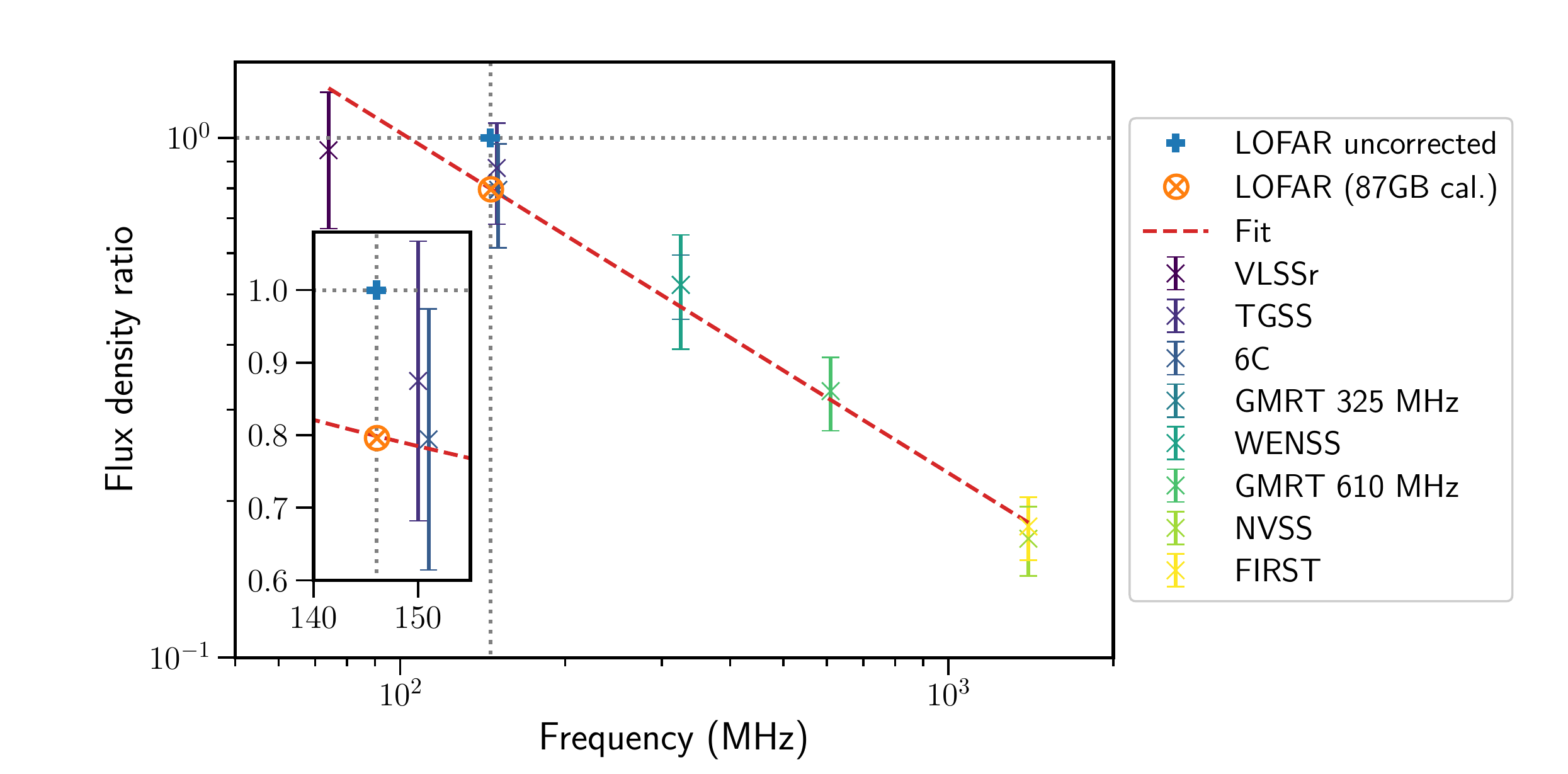}
   \caption{Calibration of the ELAIS-N1 flux density scale using different 
methods. The default \textsc{DDF-pipeline} scale is set to unity and marked with 
a blue cross. The scaling factor obtained with the 87GB calibrator is marked 
with an encircled orange cross. The flux density ratios with respect to other 
surveys in the literature and their errors are shown in different colours and 
the fitting line is shown as a red dashed line. The inset shows a zoom view of 
the area close to the 146 MHz frequency of the observation.}
   \label{fig:fluxscale}
\end{figure*}

After the filtering was applied, the ratio between the LOFAR and the survey flux 
density was computed and the results are shown in Fig.~\ref{fig:fluxscale}. Some 
surveys like TGSS or WENSS present some region-dependent issues with the flux 
density scale \citep[e.g.][]{Murphy2017}. Hence, we took into account the flux 
density uncertainty associated to a survey by adding it in quadrature to the 
error of the median of the flux density ratios. The ratio obtained for the 87GB 
calibrator is also shown in the figure. An Orthogonal Distance Regression 
\citep[ODR;][]{ODR} fit that takes into account the uncertainties in the ratios 
was fitted and used to estimate the flux density scale ratio at 146~MHz. We 
obtained a value of $0.799^{+0.052}_{-0.049}$. A fit with a second order 
polynomial gives a value of $0.795\pm0.024$ and it is favoured using the 
Bayesian information criterion \citep[BIC;][]{Schwarz1978} but not the Akaike 
information criterion \citep[AIC;][]{Akaike1998}. The fitted 146~MHz flux 
density scaling factor points lie very close to that estimated using the 87GB 
calibrator data, suggesting that the \textsc{DDF-pipeline} flux density scale is 
over-estimated by $\sim25$ per cent.

Finally, for practical reasons, we used the numerical value obtained from the 
87GB calibrator as the final value to correct the LOFAR flux density scale 
(0.796). This value is well within the confidence interval estimated with the 
linear fit with a difference of only 0.5 per cent in value (less than one tenth 
of the estimated uncertainty). We also note a likely uncertainty of around 6.5 
per cent in the flux density scale. However, we note that the uncertainty in the 
flux density scale is not included in the final quoted flux density 
uncertainties for each source. Uncertainty for individual sources can be 
affected by several factors and some of these factors will be studied in 
Sect.~\ref{sec:variability}.

The flux density scale corrections for the Lockman Hole and Bo{\"o}tes deep 
fields presented in \citetalias{Tasse2020} were also estimated using the same 
method. The final values are $0.920^{+0.041}_{-0.039}$ for the Lockman Hole and 
$0.859^{+0.036}_{-0.034}$ for the Bo{\"o}tes field. Their derivation is shown in 
Appendix~\ref{sec:LHBoo}.

\subsubsection{Flux scale of individual dataset images}
\label{sec:flux_dss}

\begin{figure*}[htbp]
   \centering
\includegraphics[width=\linewidth]{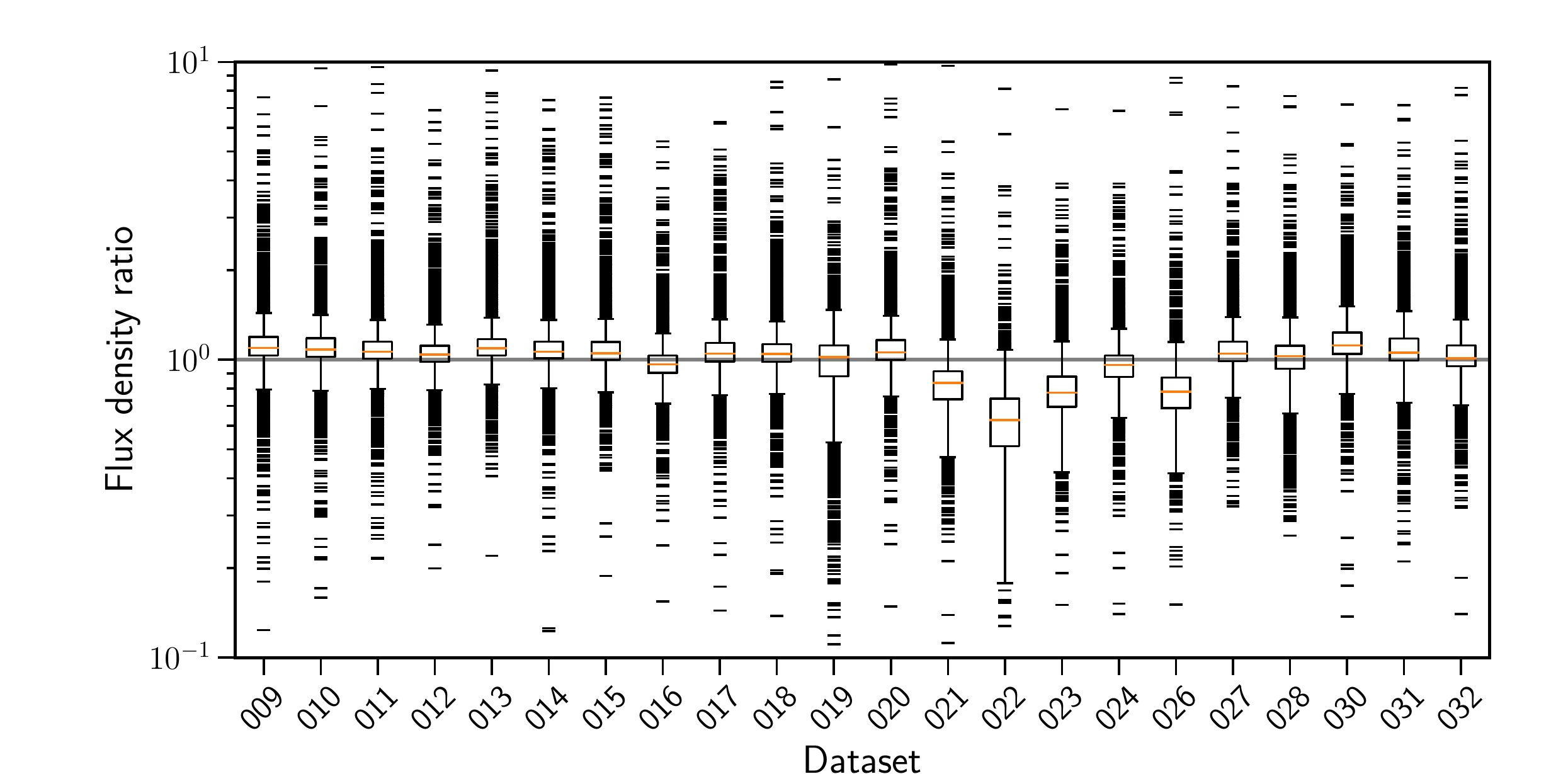}
   \caption{Distribution of the flux density ratio of the sources in common 
between the individual datasets and the final deep image. The distributions of 
the datasets are symmetrical in general in flux density scale but their median 
is systematically different from the expected value of 1. 
Fig.~\ref{fig:noise_flux_ratio} shows that the offsets are correlated with the 
noise level of the datasets.}
   \label{fig:dataset_var}
\end{figure*}

\begin{figure}
   \centering
\includegraphics[width=\linewidth]{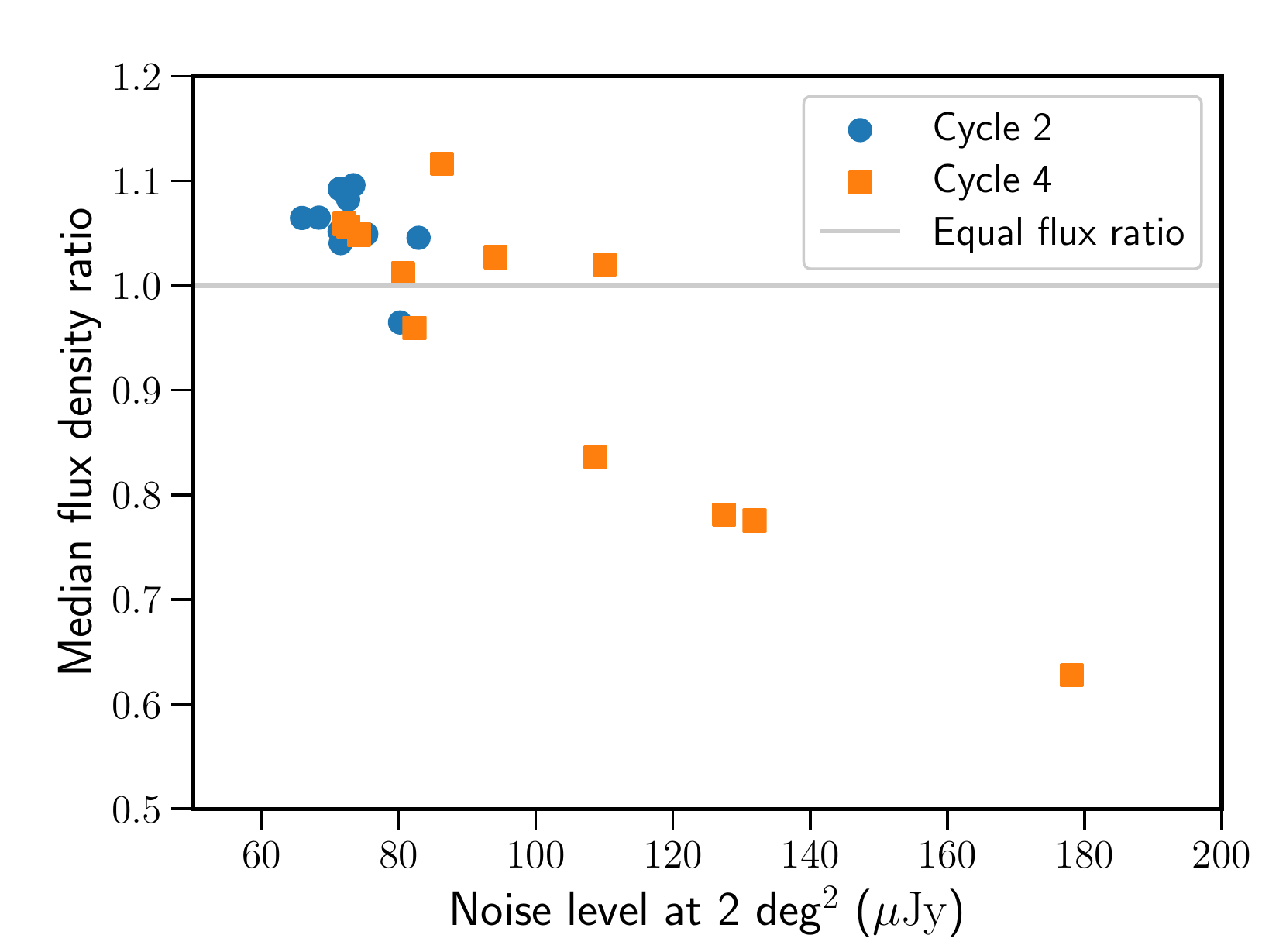}
   \caption{Relation between the deviation of the flux density ratio (observed 
over expected flux density) from the expected value and the noise level of each 
individual dataset. The vertical axis represents the median value of the flux 
density ratio distribution. The horizontal axis shows the noise level of the 
noise-area cumulative distribution at 2 square degrees. This value is a proxy 
for the calibration and data quality. Datasets of Cycle 2 are marked as blue 
circles and Cycle 4 as orange squares. The flux density ratios cluster around a 
line with value 1 (grey horizontal line) as expected but there is a clear 
inverse relation between flux density ratio and noise level.}
   \label{fig:noise_flux_ratio}
\end{figure}

The \textsc{DDF-pipeline} output flux density values for the individual dataset 
catalogues were pre-scaled by the same correction factor derived above (0.796). 
To investigate the magnitude of the residual corrections, we cross-matched the 
sources extracted from each of the individual datasets to those of the deep 
image catalogue and compared the relative flux densities of the matches. In 
Fig.~\ref{fig:dataset_var} the distribution of flux density ratios for the 
cross-matched sources is shown. The medians of the distributions of flux density 
ratios are close to one which is expected after the pre-scaling factor was 
applied. Nevertheless, for many datasets, there is a non negligible deviation 
from this value. These medians and the final scaling factors obtained from 
combining these values with the pre-scaling factor are summarized in 
Table~\ref{tab:individual_correction}. In Fig.~\ref{fig:noise_flux_ratio} we 
show the relation of these medians with the overall noise level of the dataset. 
To estimate the latter in a robust manner, we use the RMS noise level at an 
accumulated area of 2 square degrees using the cumulative distribution of area 
with respect to the noise level. This value is a proxy for the quality of the 
calibration and data. From the figure it is clear that the flux densities are 
systematically underestimated as the quality of calibration and data gets worse. 
This effect will be further studied in Sect.~\ref{sec:variability}.

\begin{table}
\centering
\caption{Flux scale correction factors for the individual datasets. These are 
the scaling factors determined by comparing the flux density of the individual 
dataset sources to that of the final deep image and by using the cross-matching 
to external surveys. The columns are the following: i) Dataset code, ii) median 
of the flux density after the pre-scaling (observed versus corrected), iii) 
total scaling factor selected, iv) scaling factor obtained using the fit to 
external surveys.}
\label{tab:individual_correction}
\begin{tabular}{lccc}
\hline
 Dataset & flux density & scaling & survey fit \\
   & ratio median & factor &  \\
\hline
\hline
009 & 1.096 & 0.726 & $0.744 \pm 0.026$\\
010 & 1.082 & 0.736 & $0.761 \pm 0.024$\\
011 & 1.065 & 0.747 & $0.755 \pm 0.024$\\
012 & 1.040 & 0.765 & $0.774 \pm 0.025$\\
013 & 1.092 & 0.729 & $0.751 \pm 0.023$\\
014 & 1.064 & 0.748 & $0.747 \pm 0.026$\\
015 & 1.052 & 0.757 & $0.771 \pm 0.028$\\
016 & 0.965 & 0.825 & $0.853 \pm 0.024$\\
017 & 1.049 & 0.759 & $0.766 \pm 0.030$\\
018 & 1.046 & 0.761 & $0.781 \pm 0.026$\\
019 & 1.020 & 0.780 & $0.828 \pm 0.026$\\
020 & 1.059 & 0.752 & $0.780 \pm 0.025$\\
021 & 0.836 & 0.952 & $0.879 \pm 0.030$\\
022 & 0.628 & 1.268 & $1.003 \pm 0.038$\\
023 & 0.775 & 1.027 & $1.014 \pm 0.028$\\
024 & 0.959 & 0.830 & $0.839 \pm 0.024$\\
026 & 0.781 & 1.019 & $0.881 \pm 0.032$\\
027 & 1.049 & 0.759 & $0.790 \pm 0.025$\\
028 & 1.027 & 0.775 & $0.765 \pm 0.028$\\
030 & 1.116 & 0.713 & $0.735 \pm 0.031$\\
031 & 1.056 & 0.753 & $0.811 \pm 0.025$\\
032 & 1.012 & 0.787 & $0.842 \pm 0.027$\\
\hline
\end{tabular}
\end{table}

We also applied the method presented previously which cross-matches to the 
external surveys. The values obtained are also presented in 
Table~\ref{tab:individual_correction}. They are usually similar but differ in 
the datasets that are more noisy. We favour the method of cross-matching to the 
final deep image as the number of cross-matched sources to obtain the estimate 
is substantially higher. 

\subsubsection{Flux scale of band images}
\label{sec:flux_bands}

The scaling factors for the bands shown in Table~\ref{tab:bands} (0, 1, 2, X, Y, 
and Z) were estimated using the method of cross-matching to the external 
surveys. The results are shown in Table~\ref{tab:band_correction}. These scaling 
factors are critical to obtain reliable intra-band spectral indices (see 
Sect.~\ref{sec:alpha}). 

To check the robustness of the method, we selected the 19264 sources that: a) 
were cross-matched between the three bands X, Y and Z within a 1 arcsec radius, 
b) had a S/N of more than 15, and c) had a major axis of less than 15 arcsecs. 
A fit to a simple Bayesian model can be used to estimate the additional scaling 
corrections that must be applied to the band scalings to get a given mean 
intra-band spectral index. We used \textsc{PyMC3} \citep{Salvatier2016} to fit 
this model which was configured to obtain a final spectral index of $\alpha = 
0.63$ \citep{Sabater2019}. The corrections found were applied and this produced 
scaling factors that were similar within the error to those obtained using the 
method of fitting to external surveys (see Table~\ref{tab:band_correction}).

\begin{table}
\centering
\caption{Flux scale factors for the different bands. The scale factors 
found for the band catalogues using the cross-match to external surveys (which 
is favoured as explained in the text), and a Bayesian model tuned to produce a 
final spectral index of $\alpha = 0.63$.}
\label{tab:band_correction}
\begin{tabular}{lcc}
\hline
 Band & Surveys fit  & PyMC3 \\
\hline
\hline
X & $0.882\pm0.032$ &  $0.86\pm0.16$ \\
Y & $0.752\pm0.028$ &  $0.77\pm0.33$ \\
Z & $0.741\pm0.025$ &  $0.75\pm0.27$ \\
0 & $0.877\pm0.032$ &  $0.85\pm0.16$ \\
1 & $0.768\pm0.028$ &  $0.77\pm0.34$ \\
2 & $0.764\pm0.025$ &  $0.77\pm0.25$ \\
\hline
\end{tabular}
\end{table}


\section{Final data products and catalogues}
\label{sec:final_catalogue}

To produce the final radio catalogues the image was scaled by the scaling factor 
determined in the previous section (0.796 for the deep image) and 
\textsc{PyBDSF} was run again using the same parameters (see 
Table~\ref{tab:pybdsf_params}). A catalogue of sources and a catalogue of 
individual Gaussians are produced. The columns of the catalogues are those of a 
typical \textsc{PyBDSF} default output including the position, integrated and 
peak flux density, structural parameters (raw and deconvolved) and their 
estimated errors \citep{Mohan2015}. The radio catalogue produced in this way, 
and presented in this paper, is in raw state and the generation of a final 
source catalogue requires further processing to consider blended sources, 
artefacts, and the merging of some \textsc{PyBDSF} components into a single 
source. All of that work, and the cross-correlation with multiwavelength 
catalogues is presented in \citetalias{Kondapally2020}. 

\begin{figure}[htbp]
   \centering
\includegraphics[width=\linewidth]{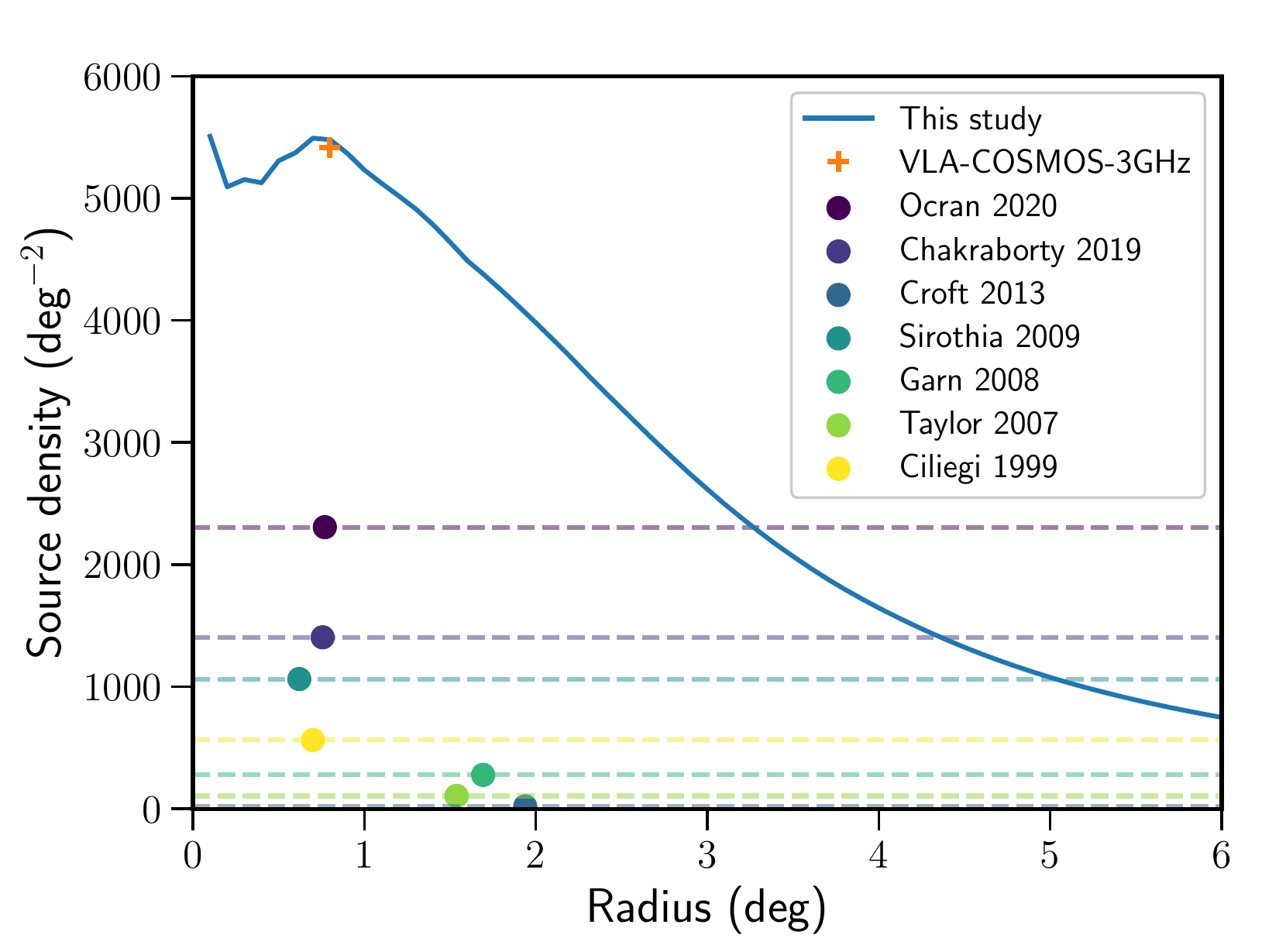}
   \caption{Comparison of the sky density of radio sources in our work with 
those quoted by other radio ELAIS-N1 surveys. The solid blue line shows the 
total density of sources at each radius for our study. The points show the 
density of sources quoted in the literature at the corresponding radius to the 
area covered by those surveys. The density is extended with dashed lines to give 
an idea of the radius out to which a similar average source density would be 
achieved with our data. We also show as a comparison the estimated sky density 
of VLA-COSMOS-3GHz \citep{Smolcic2017} which reaches a comparable source density 
in the total 2 square degree area that it covers.}
   \label{fig:comparison_density}
\end{figure}

The deep radio catalogue contains 84862 sources composed of 96039 Gaussians. 
There are 78885 sources composed of a single Gaussian and 5951 multicomponent 
sources that correspond to 17128 Gaussians. The number of sources within 1, 2, 
and 3 degrees from the centre are 16435, 50026, and 74127 respectively. 
Fig.~\ref{fig:comparison_density} shows the total sky density of sources at 
different radii. It also compares the sky density to that obtained for previous 
targeted surveys of ELAIS-N1. It shows that the LOFAR data are much deeper as 
well as wider and at the same time it typically has a better resolution.

We also extracted catalogues for the 6 spectral bands shown in 
Table~\ref{tab:bands}. Finally, we extracted catalogues for the individual 
datasets used in the deep image. All these catalogues were extracted with 
\textsc{PyBDSF} using the same parameters (see Table~\ref{tab:pybdsf_params}). 
The spectral band and individual dataset catalogues are offered with the flux 
density scaling factors shown in Table~\ref{tab:individual_correction} and 
\ref{tab:band_correction} applied.


\section{Discussion}
\label{sec:discussion}

\subsection{Noise levels and source confusion}
\label{sec:noise}

\begin{figure*}[htbp]
   \centering
\includegraphics[width=\linewidth]{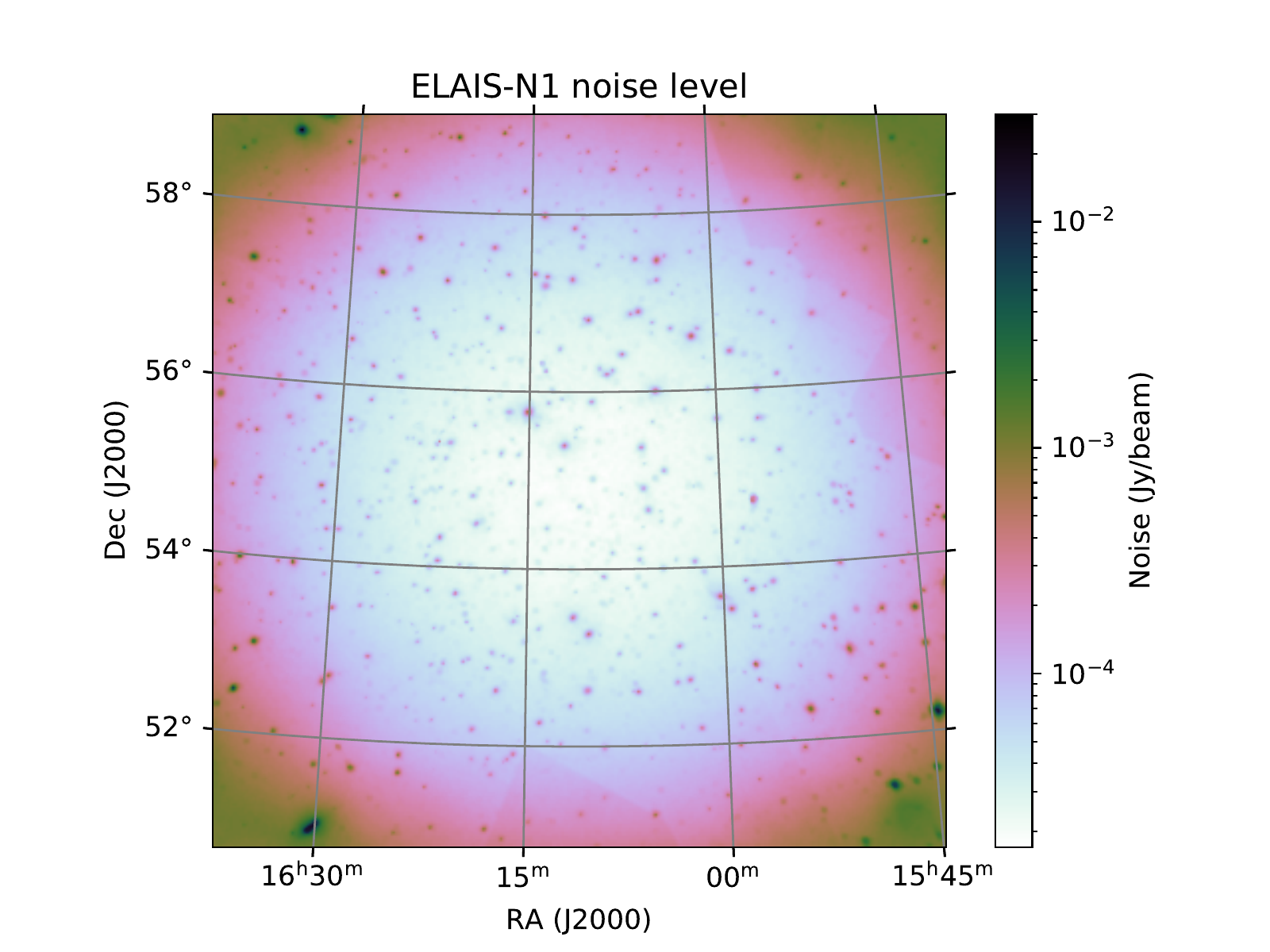}
   \caption{RMS noise level in the ELAIS-N1 field. The minimum value is close to 
the centre at a level of \mbox{17.1 $\mu$Jy beam$^{-1}$}. The effect of the 
facet division used for the calibration is only visible in areas very far from 
the centre. The small calibration errors around very bright sources have an 
adverse effect in the dynamic range of the nearby region.}
   \label{fig:noise}
\end{figure*}

\begin{figure}[htbp]
   \centering
\includegraphics[width=\linewidth]{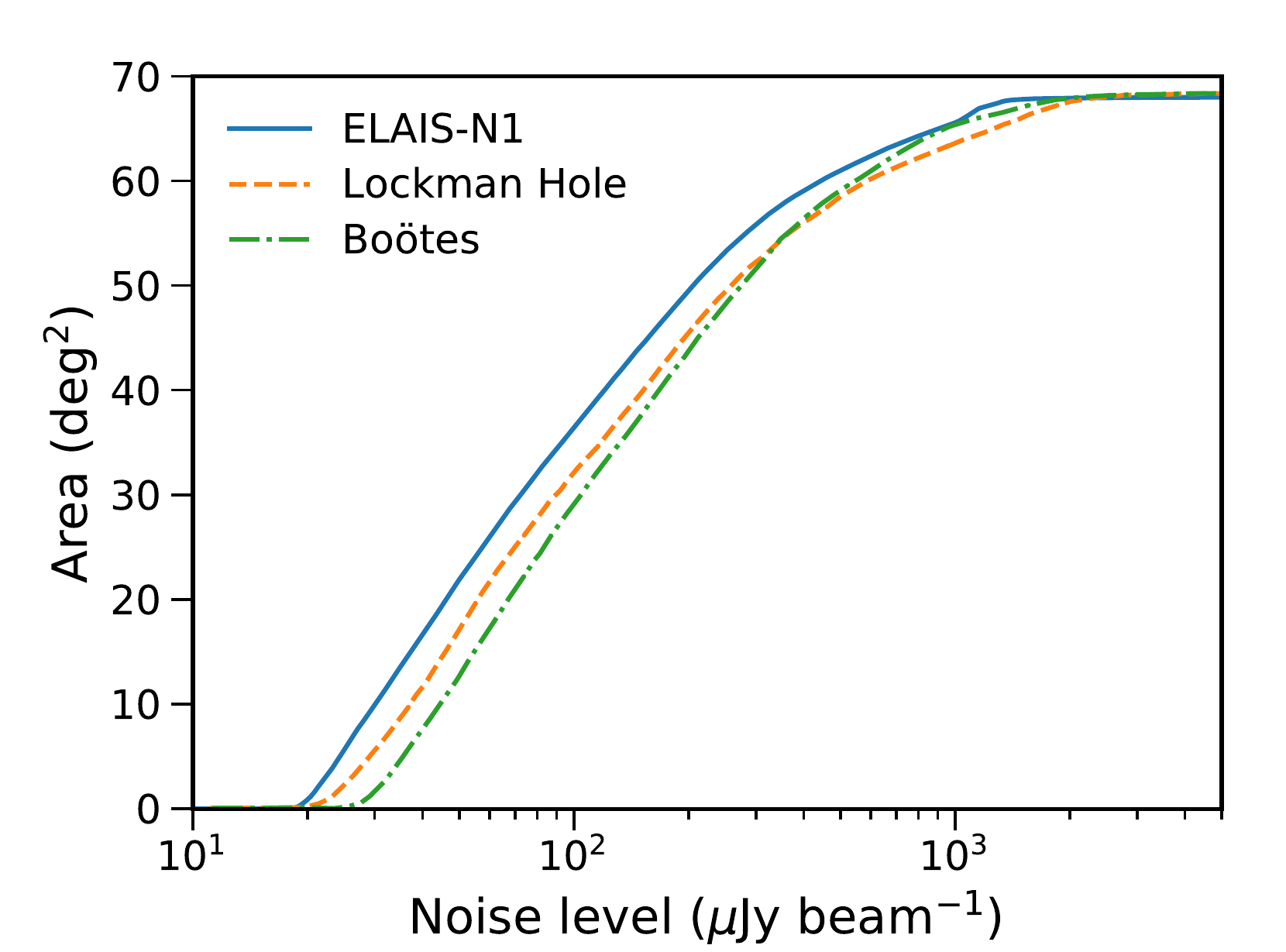}
   \caption{Area coverage with RMS noise equal to or better than a given value. 
The distribution for ELAIS-N1 (solid blue line) is compared to that of the other 
two LoTSS deep fields, the Lockman Hole (dash-dot green line) and Bo{\"o}tes 
(dashed orange line), observed using similar techniques as presented in 
\citetalias{Tasse2020}.}
   \label{fig:area_noise}
\end{figure}

The RMS noise level of the deep image (see Fig.~\ref{fig:noise}) reaches a 
minimum value of \mbox{17.1 $\mu$Jy beam$^{-1}$}. The median value in the 
$\approx 7$\,deg$^2$ area covered by the added value datasets 
\citepalias{Kondapally2020} is \mbox{22.9 $\mu$Jy beam$^{-1}$}. The cumulative 
distribution of area with respect to the noise level is shown in 
Fig.~\ref{fig:area_noise}. The distribution is also compared with those of the 
Bo{\"o}tes and the Lockman Hole deep fields \citepalias{Tasse2020}. They present 
slightly higher noise levels as they were observed for a shorter period of time 
($\sim100$ and $\sim112$ hours respectively) and have smaller bandwidths (48 MHz 
each). In the case of Bo{\"o}tes it is also located at a significantly lower 
declination ($\sim34$\degr) and additional noise produced by the projection of 
the station beams is added.

Factors that would limit the depth of the images as additional data are combined 
include confusion noise from faint sources below the flux density limit of the 
survey, or the growing importance of residual calibration errors. The latter 
leads to the reduction of the dynamic range around the brightest sources (see 
Fig.~\ref{fig:noise}),  but can have also wider effects due to correlated 
sidelobe structures around less bright sources. 

\begin{figure*}[htbp]
   \centering
\includegraphics[width=\linewidth]{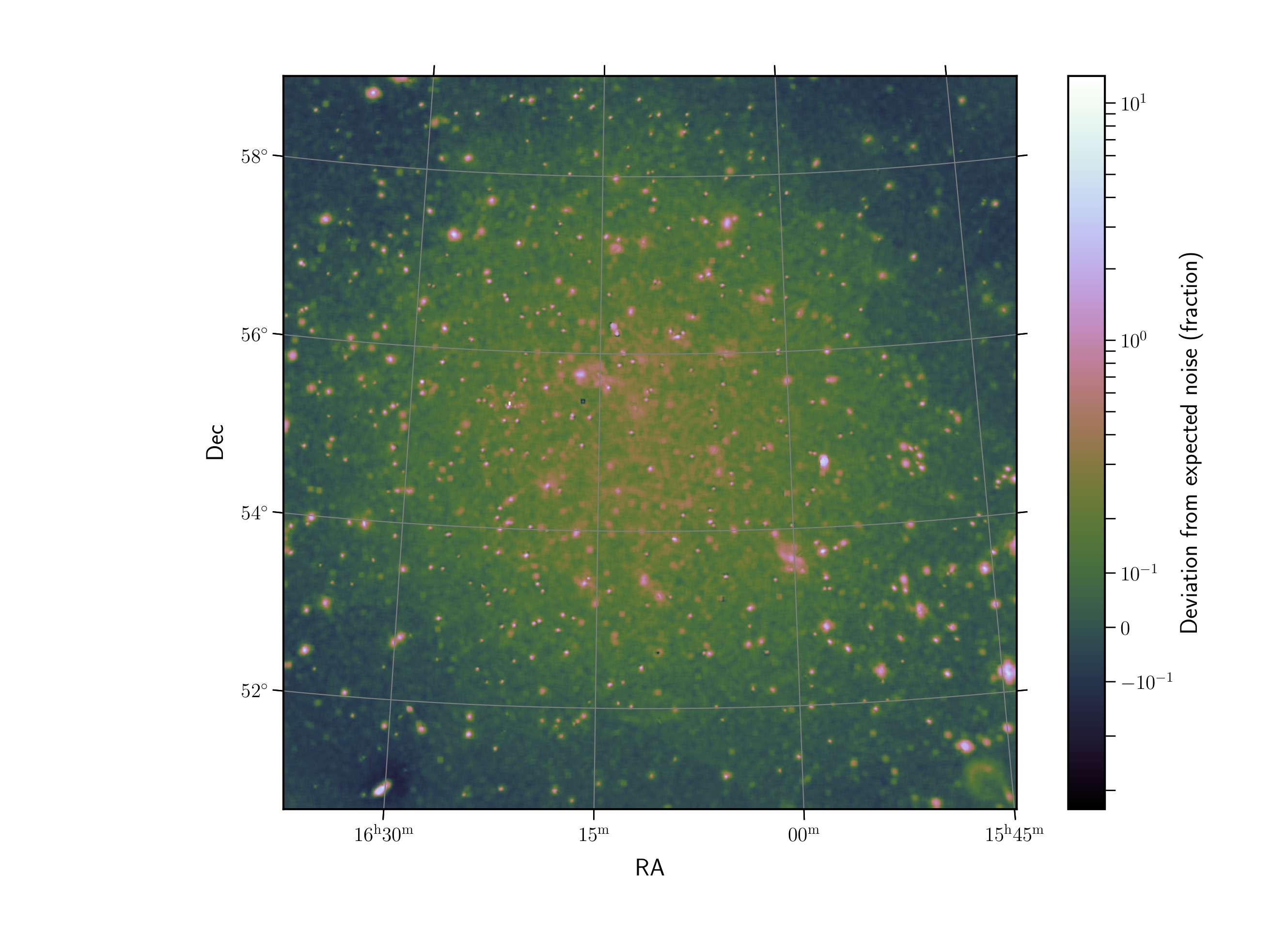}
   \caption{Noise deviation in the combined image with respect to that predicted 
from the combination of the individual dataset images. In the areas with high 
deviations from the expected value these are limited because the dynamic range 
of the deep image is reduced due to residual errors in the calibration around 
bright sources.}
   \label{fig:noise_dev}
\end{figure*}

\begin{figure}[htbp]
   \centering
\includegraphics[width=\linewidth]{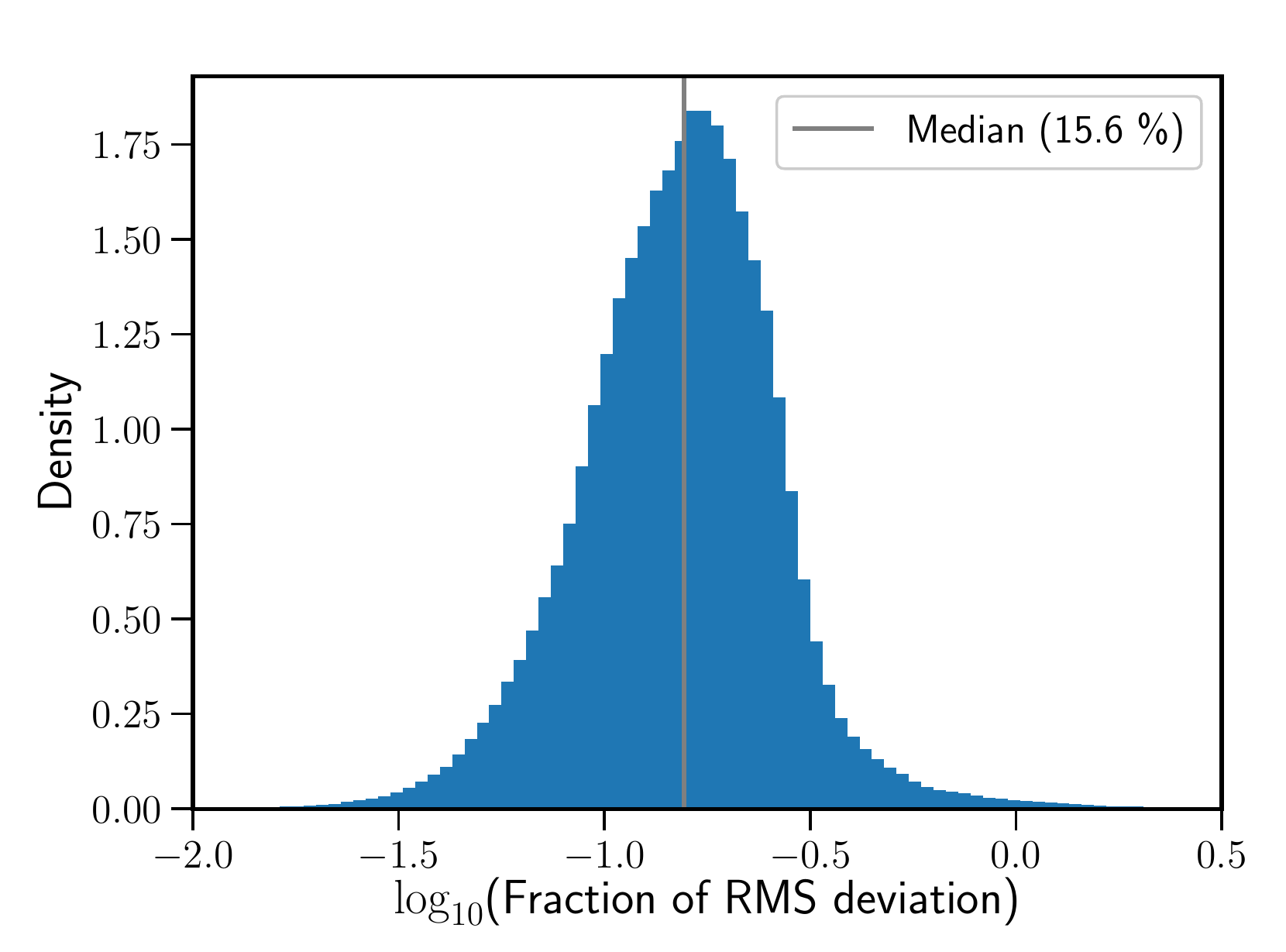}
   \caption{Histogram of the noise deviation in the combined image with respect 
to that predicted from the individual images, assuming that the noise in each is 
uncorrelated. The histogram is computed in the area closer than 3 degrees to the 
centre. The median value of the distribution (solid grey line) is a 15.6 per 
cent increase in noise in the deep image.}
   \label{fig:noise_dev_hist}
\end{figure}

To investigate the magnitude of these effects, we compared the noise level of 
the final deep image with that predicted by theoretical combination of the noise 
from the individual images (assuming that the noise in each image is completely 
uncorrelated). If there is negligible confusion noise and no errors in the 
calibration, the values should be similar. The comparison between the predicted 
and obtained noise levels are shown in Fig.~\ref{fig:noise_dev}. A histogram for 
the values in the inner 3 degrees region is shown in 
Fig.~\ref{fig:noise_dev_hist}. There are some regions around bright sources in 
which the dynamic range limitations are clearly visible but for most of the area 
the observed noise level in the final image is less than 16 per cent higher than 
the ideally predicted value. This suggests that any increase in noise due to 
imperfect calibration is relatively small. Interestingly, there are some areas 
at the edge where the obtained noise level is better than that predicted from 
the individual images. This may arise from several factors like, for example, a 
reduction in the noise produced by faint sources which were not able to be 
properly deconvolved in the individual datasets, but may also reflect the 
improvement in the performance of the third generation calibration algorithms as 
the amount of data fed increases \citepalias{Tasse2020}. 

For the increased noise in the central regions to be caused entirely by source 
confusion, a confusion noise of 11$\mu$Jy~beam$^{-1}$ would be required. For 
shallower LoTSS observations, \citet{Shimwell2019} estimated a confusion level 
for LOFAR 150~MHz observations of $14 \mu$Jy~beam$^{-1}$, but with a wide 
uncertainty range depending upon the slope of the faint number counts. 
\citet{Condon2012} estimated the confusion noise at higher frequencies using 
ultra-deep 1.4 and 3~GHz data, and provide an equation (their Eq.~27) for 
conversion of these to other frequencies: $\sigma_c = 1.2 (\nu / 
3.02\mathrm{GHz})^{-\alpha} (\theta / 8\mathrm{arcsec})^{10/3} 
\mu$Jy~beam$^{-1}$, where $\nu$ is the frequency of observation and $\theta$ is 
the synthesized beam size. For the ELAIS-N1 data ($\nu = 146$~MHz; 
$\theta=6$~arcsec), taking $\alpha=0.63$ as a median spectral index 
\citep{Sabater2019}, this gives an estimate for the confusion noise of $\sigma_c 
\approx 3 \mu$Jy~beam$^{-1}$. However, this estimate involves a long 
extrapolation in frequency, which may not be appropriate.

The confusion noise level can be directly estimated from the source count 
distribution \citep[e.g.][]{Condon2012}: $$\sigma_c = 
\left(\frac{q^{3-\gamma}}{3-\gamma} 
\right)^{1/(\gamma-1)}(k\Omega_e)^{1/(\gamma-1)},$$ where $q$ is the S/N 
threshold of the catalogue, $k$ and $\gamma$ are the normalization and slope of 
a power-law fit to the source counts at low flux densities, $N(S) = k 
S^{-\gamma}$, and $\Omega_e$ is the effective beam solid angle, given by 
$\Omega_e = \pi \theta^2 / (4\ln(2) (\gamma-1))$.  The values of $k$ and 
$\gamma$ are very sensitive to incompleteness corrections at the faint end: 
analysis of these is beyond the scope of this paper, but is instead investigated 
in the accompanying paper of \citet{Mandal2020}. In ELAIS-N1, the 
Euclidean-normalized source counts begin to turn down below about $S_{\rm 150 
MHz} \approx 400\mu$Jy; both theoretical models and high-frequency observations 
suggest that this will turn over to a slope of $\gamma \approx 1.5-1.8$.  
Normalizing to the observed counts at 200$\mu$Jy ($N(S) S^{5/2} \approx 
30$\,sr$^{-1}$Jy$^{1.5}$) provides an estimate for the confusion noise of 
$\approx 8 \mu$Jy~beam$^{-1}$ for $\gamma = 1.7$. The precise confusion noise, 
however, remains sensitive to the choice of $\gamma$, which the data do not yet 
constrain to sufficient accuracy. We conclude that confusion noise is likely to 
account for a significant fraction of the median increase in noise in the 
central regions compared to the predicted value, but probably does not account 
for all of it.

\subsection{Estimation of source flux density variation}
\label{sec:variability}

The imaging of the individual datasets allows to perform an analysis of the 
empirical variation between observations of the source flux density. Calibration 
effects should be disentangled from intrinsic variability and the first step was 
to bring all the datasets to the same flux density scale frame using their 
individual correction factors as presented in Sect.~\ref{sec:flux_dss}. 

To measure the variations of the flux density and determine their origins, we 
studied the sources that were detected in common in all the individual images. 
There are 5504 sources cross-matched and detected in all the 22 individual 
snapshots. For all these sources, we computed the relative flux density of each 
individual observation with respect to the flux density estimated from the deep 
image. The values of these relative fluxes have a mean close to 1 and on a log 
scale their distribution is in general symmetrical on both sides of the zero 
value. We have studied the standard deviation of this distribution 
($\log_{10}(S_{i}/S_{0})$; hereafter called $\sigma$) applying different 
constraints to the data in order to find what the most important factors 
affecting it are.

\begin{figure*}[htbp]
   \centering
\includegraphics[width=0.45\linewidth]{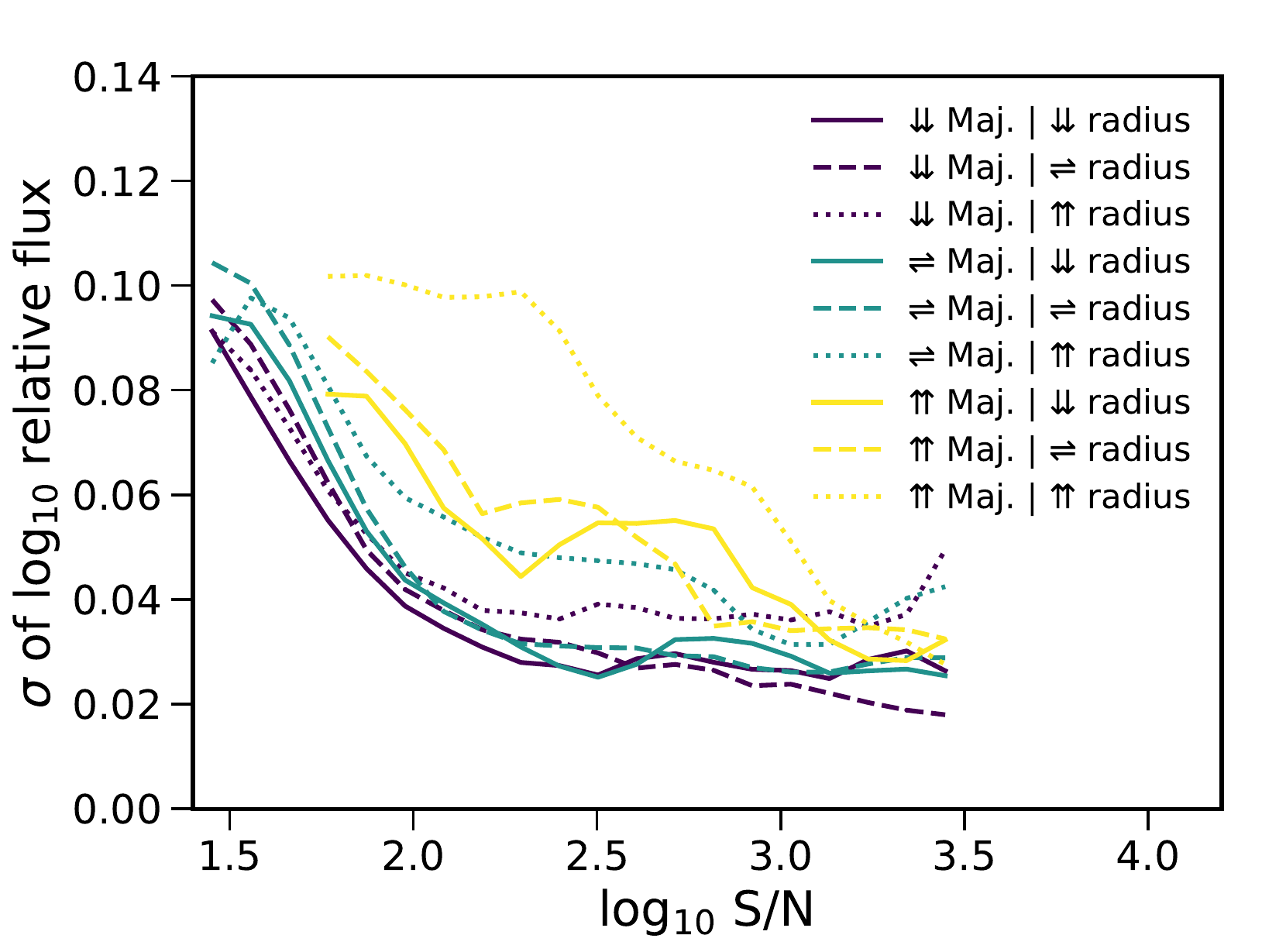}
\includegraphics[width=0.45\linewidth]{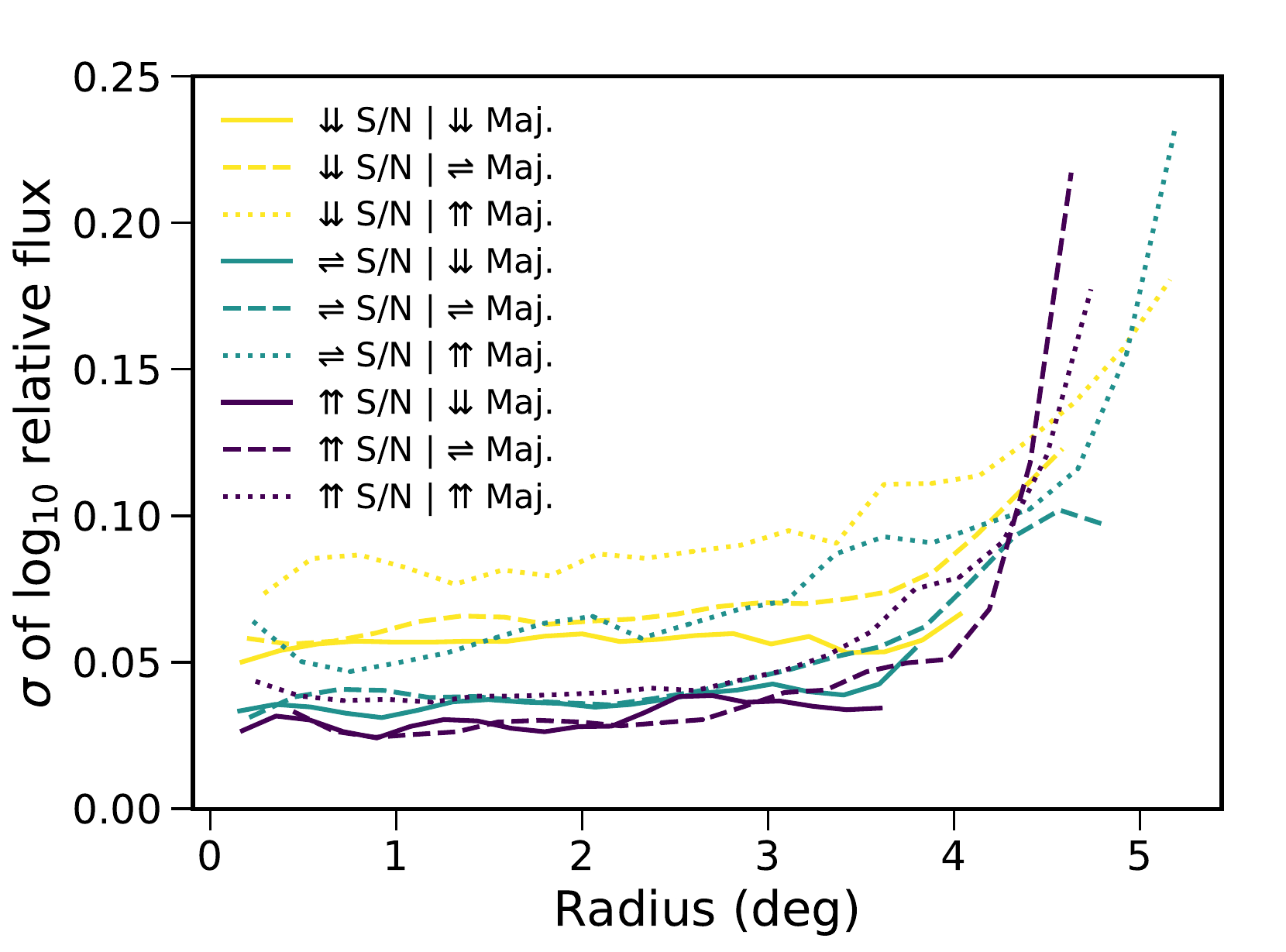}
\includegraphics[width=0.45\linewidth]{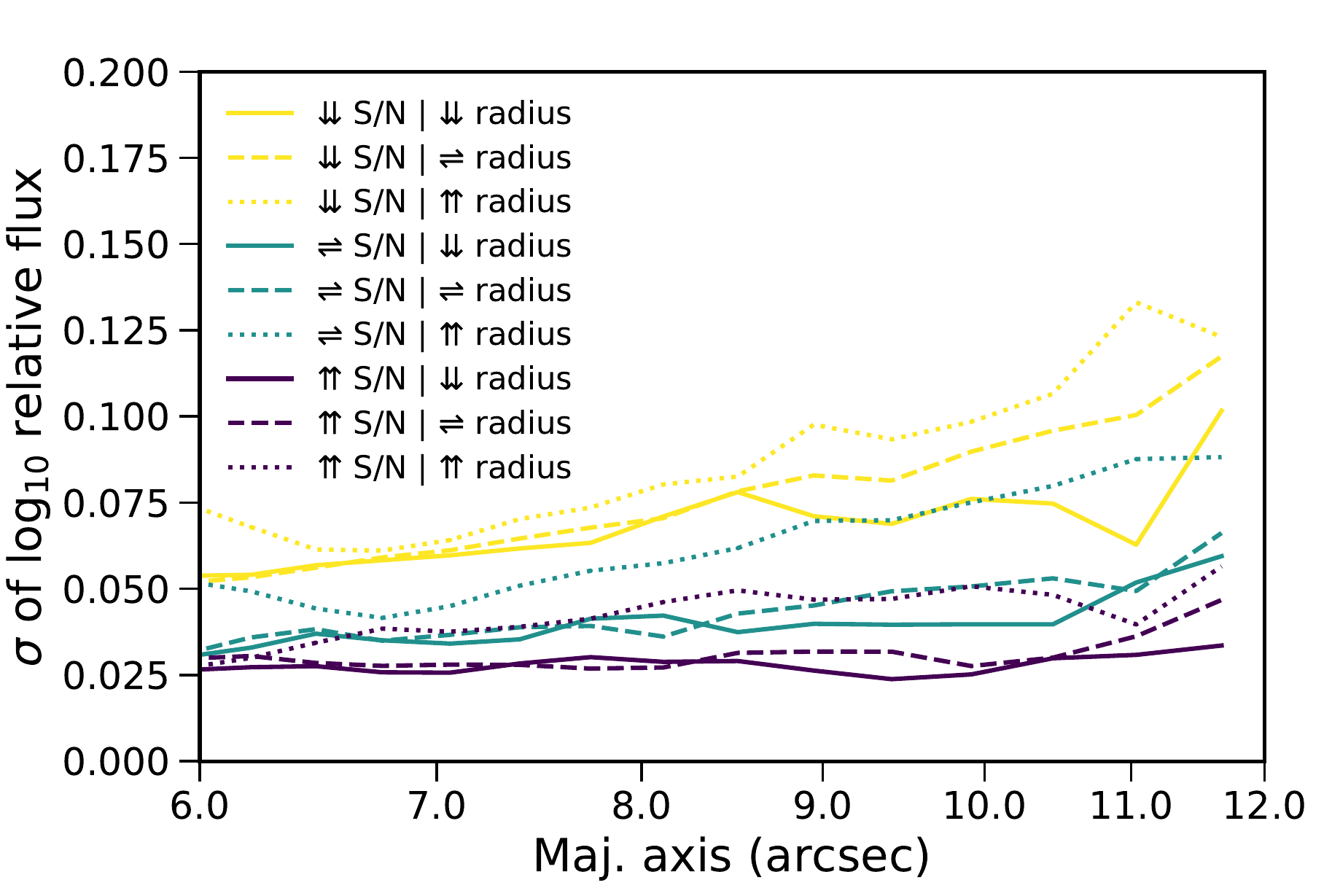}
\includegraphics[width=0.45\linewidth]{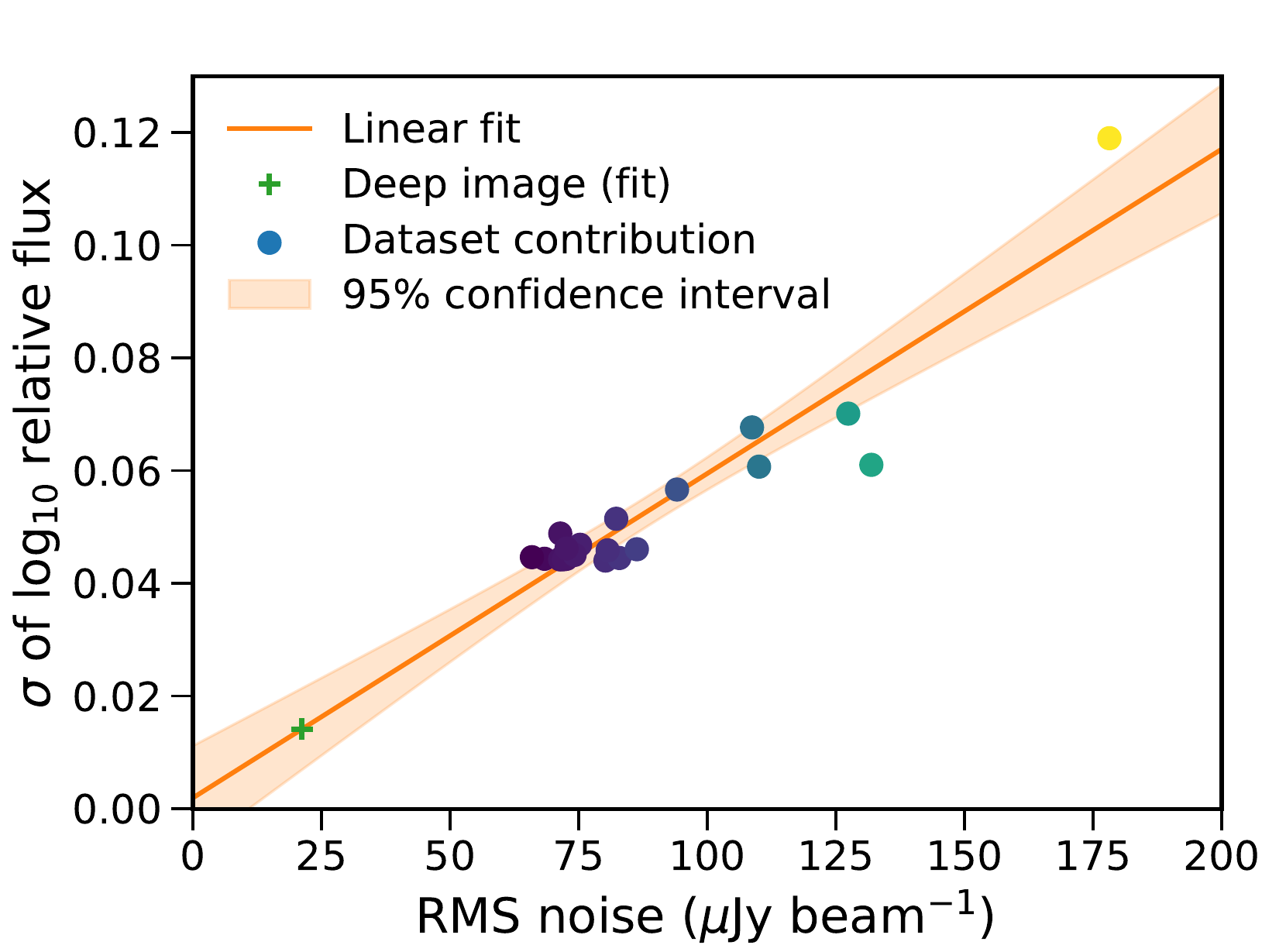}
   \caption{Standard deviation ($\sigma$) of the logarithmic relative flux 
density with respect to several parameters. The upper left panel shows the 
relation of $\sigma$ with respect to the signal-to-noise level for sub-samples 
separated in terciles of distance from the pointing centre and major axis. The 
different terciles are marked with arrow symbols in the legend: up arrows for 
the higher tercile, left-right arrows for the middle tercile, and down arrows 
for the lower one. The upper right panel shows the relation of $\sigma$ with 
respect to the distance to the centre for terciles of major axis and S/N. The 
lower left panel show the relation of $\sigma$ with respect to the major axis of 
the sources for terciles of distance to the centre and S/N. The lower right 
panel shows the contribution of the calibration effects of the dataset 
($\sigma_{dataset}$, as explained in the text) with respect to the RMS noise at 
2 square degrees for the individual datasets (coloured circles). The linear fit 
to the relation is shown as a orange line and its 95 per cent confidence 
interval as a shaded orange band. The extrapolated location of the final deep 
image is also marked in the diagram as a green cross.}
   \label{fig:sigma}
\end{figure*}

We have identified four main factors that have a significant effect in the 
variation of the relative flux densities: the radial distance from the centre, 
the size of the source, the S/N of the measurement, and the overall noise level 
of the dataset image. The effect of these parameters in $\sigma$ are shown in 
Fig.~\ref{fig:sigma}. To study the effect of the radius, size, and S/N we have 
partitioned the sample in sub-samples according to terciles of their 
distribution. The thresholds are: a) in radius, 1.7 and 2.7 degrees from the 
centre; b) in size, major axes of 6.9 and 8.9 arcsecs; and c) in S/N, the values 
of 89 and 227. We studied the relation of $\sigma$ with each parameter for 
sub-samples comprising the combination of terciles of the remaining two 
parameters, while dividing the studied parameter in $\approx 20$ bins. The 
$\sigma$ was computed for each individual bin and the result smoothed with a 
Savitzky-Golay filter \citep{SavitzkyGolay1964}. 

The effect of the S/N in $\sigma$ is shown in the upper left panel of 
Fig.~\ref{fig:sigma}. As it would be expected, this is very pronounced at lower 
signal-to-noise values, decreasing rapidly as S/N increases. However, for all 
but extended sources, it then flattens out (rather than continuing to decrease 
towards zero) once a S/N of 100--300 is exceeded, at a floor value of $\sigma 
\approx 0.025$. The effect of the radius is shown in the upper right panel of 
Fig.~\ref{fig:sigma}. The $\sigma$ is independent of radius from the pointing 
centre until a radius of between 3 and 4 degrees is reached, after which it 
rises very steeply. This effect can be produced by the quality of the 
calibration degrading in facets far from the centre but also due to errors in 
the modelling of the primary beam shape. The effect of the size of the source is 
shown in the lower left panel of Fig.~\ref{fig:sigma}. In this case there is no 
visible relation for sources with high S/N and at low and mid radii. For the 
rest there is an increase in $\sigma$ with the size that depends on their S/N 
and radius. This is likely related to the \textsc{PyBDSF} extraction of sources. 
For extended sources with low S/N the \textsc{PyBDSF} fitted Gaussians can be 
different from one dataset to the next, introducing artificial variability.

Finally, the effect of the dataset can be estimated in a sample that minimizes 
the effect of the remaining parameters. We select the 240 sources with a S/N 
above 300, a distance to the centre of less than 3 degrees, and a major axis of 
less than 7 arcsecs. For each of these sources, we calculate the observed 
distribution of $\sigma$ associated with that dataset ($\sigma_{observed}$), and 
also calculate the expected distribution, based on the tabulated flux densities 
and uncertainties for each source ($\sigma_{expected}$). We then estimate the 
contribution of the dataset as the difference in quadrature between the observed 
and the expected $\sigma$ based on the individual measurements 
($\sigma_{dataset}^2 = \sigma_{observed}^2 - \sigma_{expected}^2$). In the lower 
right panel of Fig.~\ref{fig:sigma} we show the dependence of $\sigma_{dataset}$ 
on an estimation of the noise level of the dataset. The RMS noise shown in the 
horizontal axis is the noise level at 2 square degrees based on the cumulative 
distribution for each individual dataset (see Sect.~\ref{sec:noise}). This value 
is a proxy for the quality of the data and calibration. We find that the value 
of $\sigma_{dataset}$ is clearly higher in the datasets with higher RMS noise, 
indicative of poorer calibration solutions mainly due to bad ionospheric 
conditions.

It is worth noting that the levels of $\sigma_{dataset}$ are larger now than the 
levels of $\sigma$ found for these `well-behaved' sources individually (lower 
lines in the upper and lower left panels of Fig.~\ref{fig:sigma}). This is 
produced by the extra variance introduced by the fact that the ratio between the 
flux density measurement in the deep image and the average flux density measured 
in the individual dataset is in general not exactly equal to unity for 
individual sources. If we correct by this factor the values obtained are 
similar. However, we kept this term because real measurements would include it 
as well.

We fitted a linear relation to the $\sigma_{dataset}$ versus noise level data 
and obtained $\sigma_{dataset} = 0.00194 x + 0.00058$ where $x$ is the RMS noise 
level at 2 square degrees. An extrapolation of this equation down to the noise 
level of the deep image (taking into account the confidence interval of the 
prediction) suggests a value of $\sigma_{dataset}$ of $0.014 \pm 0.004$ in the 
deep data, which is equivalent to $3.3 \pm 0.8$ per cent of the flux density 
ratio. This indicates that, even for compact central sources with the higher 
signal-to-noise ratios, there is likely to be an extra component in the flux 
density uncertainty of this magnitude. This possible error was not included in 
the final catalogues but must be taken into account if this order of precision 
is required. 

Once the magnitude of the flux density error is determined it is possible to 
robustly identify the subset of sources that present significant intrinsic 
variability. A full investigation of variable sources in the dataset, using 
advanced techniques complemented by careful visual inspection, will be presented 
in a companion paper (Sabater et al. 2020b).

\subsection{Extended sources}
\label{sec:extended}

It is of interest to identify which sources in the radio catalogue are extended 
and which are point-like. A particular technique to achieve this is explained by 
\citet{Franzen2015} and, for this study, we propose a further improvement to 
this method that adds some modifications based on the work of 
\citet{Shimwell2019}. If $R$ is defined as $R = \ln(S_{total}/S_{peak})$, its 
distribution should be Gaussian and centred on zero for unresolved sources 
\citep{Franzen2015}. The RMS of $R$ can be given as a combination in quadrature 
of different error terms as: $$ \sigma_{R} = 
\sqrt{\left(\frac{\sigma_{total}}{S_{total}}\right)^2 + 
\left(\frac{\sigma_{peak}}{S_{peak}}\right)^2 + C},$$ where the term $C$ 
accounts for additional errors and is empirically determined for our data. 

\begin{figure}[htbp]
   \centering
\includegraphics[width=\linewidth]{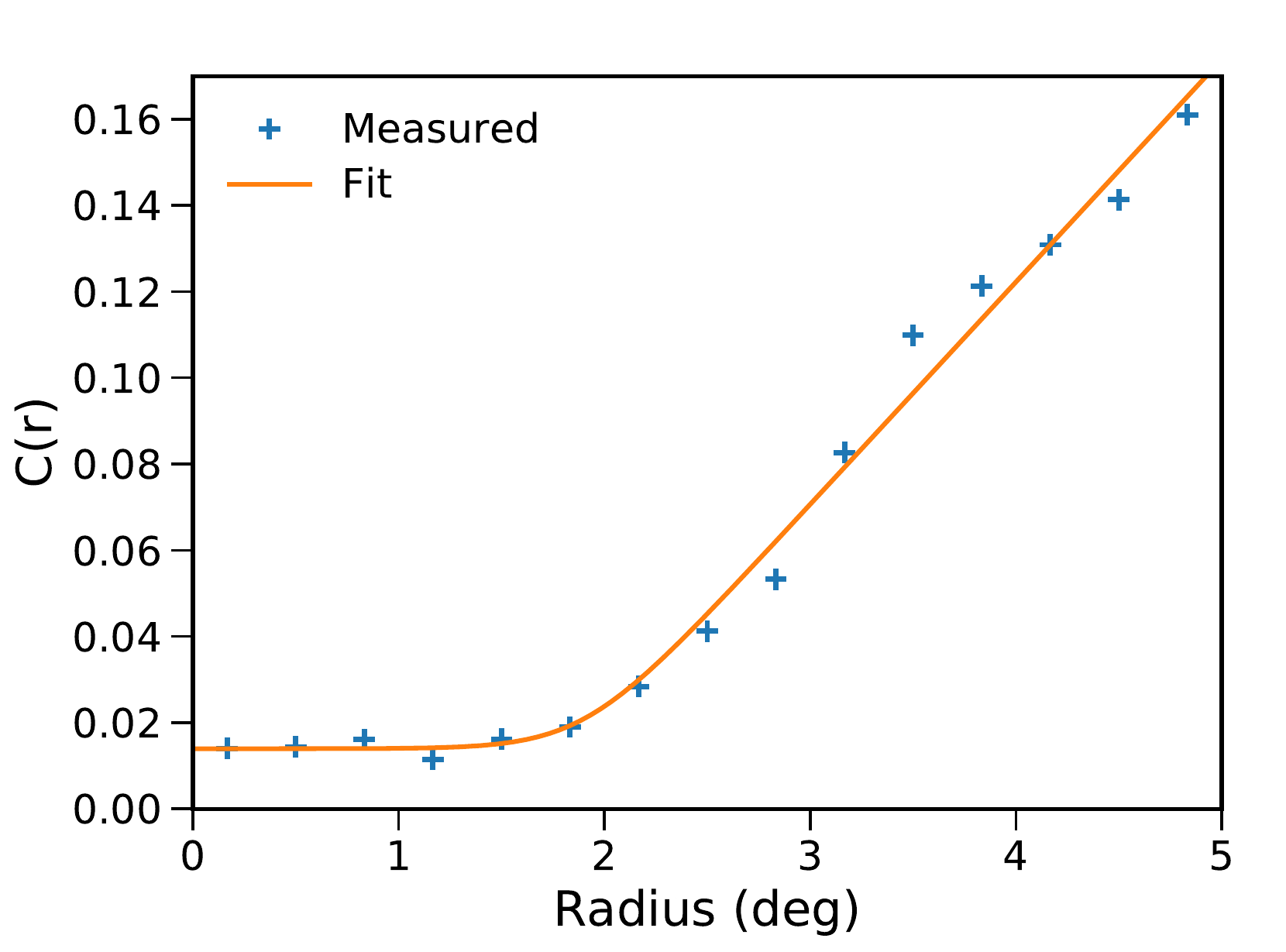}
   \caption{Dependence of the additional noise term $C$ with radius. The 
empirical values (blue crosses) are fitted with a modified softplus function 
(orange line).}
   \label{fig:cr}
\end{figure}

To empirically compute the parameter $C$, we select isolated sources as in 
\citet{Shimwell2019}, that is, sources that are: a) flagged as single sources by 
\textsc{PyBDSF} (code `S'); b) have a major axis smaller than 15 arcsecs; and c) 
have no nearest neighbours at a distance of less than 45 arcsecs. The empirical 
distribution of $C$ for these sources is strongly dependent on radius as shown 
in Fig.~\ref{fig:cr} and as explained in Sect.~\ref{sec:variability}. We fit the 
points to a modified softplus function with the following formula: $$C(r) = c_0 
+ c_1\,\ln(1 + e^{c_2 (r - c_3)}),$$ where $r$ is the radius and $(c_0, c_1, 
c_2, c_3)$ are the free parameters to be adjusted indicating respectively: the 
level of $C$ at low radii, the slope of the line at high radii, a scaling factor 
to adjust the shape of the curve, and the position of the turnover radius. The 
fit is done by minimizing the square vertical distance between the points and 
the curve using the method of \citet{Powell1964} which is implemented in 
\textsc{Scipy}. The fitted line is shown in Fig.~\ref{fig:cr} and $\sigma_{R}$ 
can be computed using $$C(r) = 0.0101 \,\ln \left(e^{5.1 (r - 1.9)} \right) + 
0.0139.$$

\begin{figure}[htbp]
   \centering
\includegraphics[width=\linewidth]{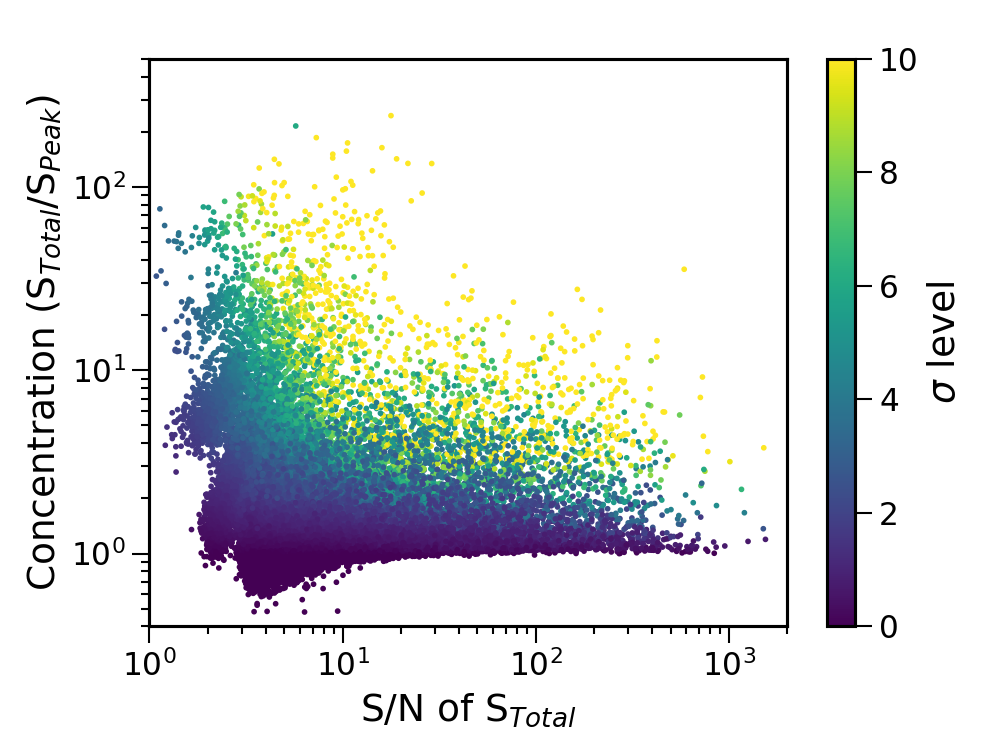}
   \caption{Diagram showing the location of extended sources. 
The concentration (total over peak flux density) is plotted with respect to the 
S/N of the total flux density. The colour scale represents the 
confidence level at which a source can be considered to be extended after also 
taking into account the radial location of the source in the field (see text 
for more details).}
   \label{fig:extended}
\end{figure}

Computing the ratio $R/\sigma_R$, we can estimate at which level the source is 
expected to be extended. In Fig.~\ref{fig:extended} we show the distribution of 
concentrations ($R$) versus S/N and the ratio in different colours. Although the 
values of the ratio depend on the individual sources, they follow clear trends 
and the shape is similar to that found in \citet{Franzen2015, Shimwell2019} and 
\citet{Chakraborty2019}. Applying a threshold of 5 sigma there are 3426 sources 
(4.0 per cent) classified as extended while a 3 sigma threshold gives 9630 
extended sources (11.3 per cent).

\subsection{Intra-band spectral index}
\label{sec:alpha}

\begin{figure}[htbp]
   \centering
\includegraphics[width=\linewidth]{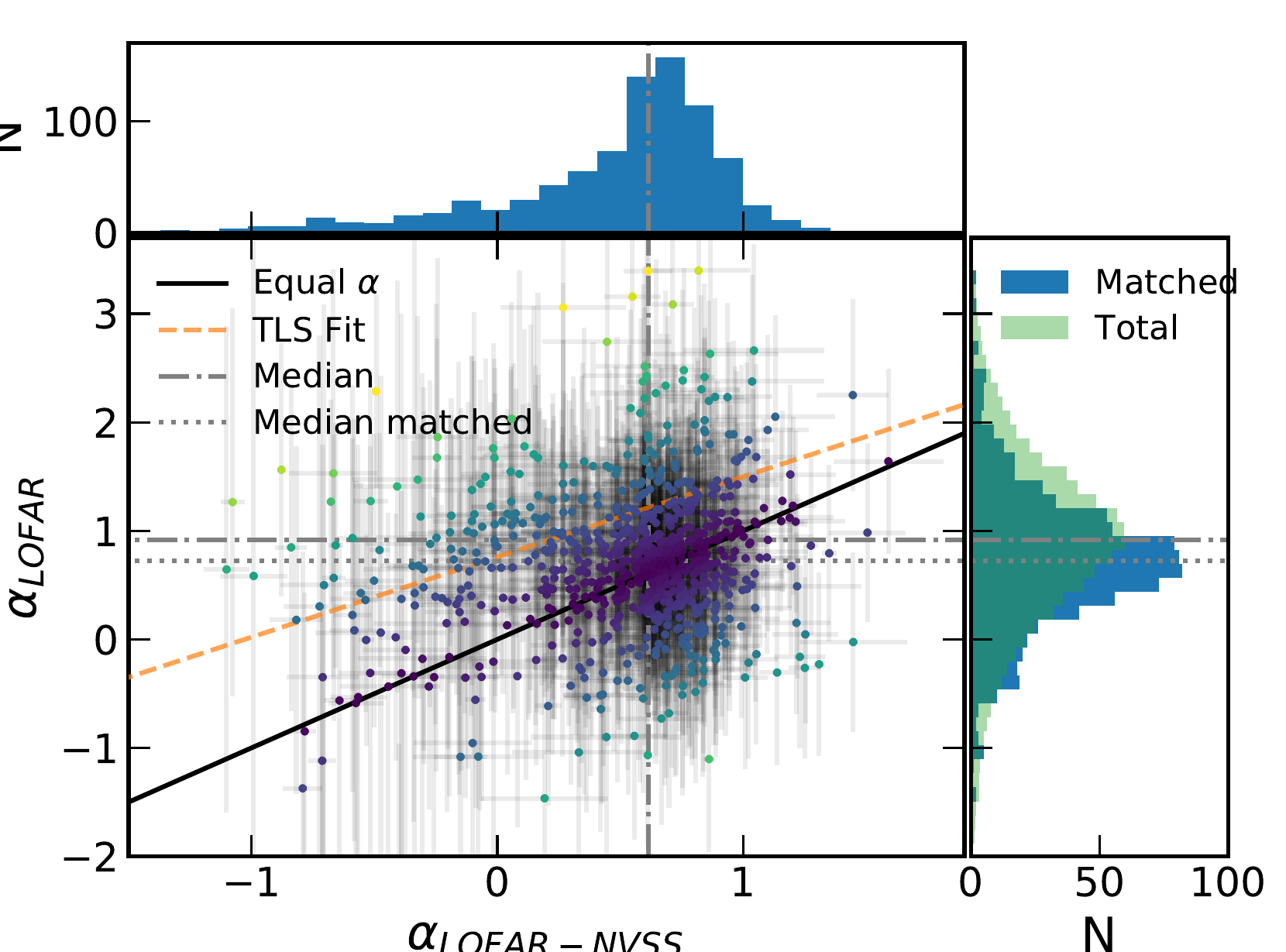}
   \caption{Comparison of the intra-band spectral index ($\alpha_{LOFAR}$) with 
the NVSS to LOFAR spectral index ($\alpha_{LOFAR-NVSS}$). The colour of the dots 
is proportional to the deviation from the (black solid) line where the spectral 
indices are equal. The errors are plotted as faint grey lines and indicate the 
95 per cent confidence interval. The median of the distributions for all the 
sources are shown as grey dash-dotted lines. The median of the distribution of 
$\alpha_{LOFAR}$ for the sources that were matched to NVSS sources is shown as a 
dotted line. The orange dashed line shows the total least squares fit to the 
distribution (see text). The upper panel shows an histogram with the 
distribution of $\alpha_{LOFAR-NVSS}$. The right panel shows the histogram with 
the distribution of $\alpha_{LOFAR}$ for the cross-matched sources (blue) and 
for the total sample over-plotted (green).}
   \label{fig:alpha}
\end{figure}

We estimated the intra-band spectral index for the sources that were matched 
between the three extended band images (X, Y and Z). The estimation uses the 
accurate calibration of the relative flux densities presented in 
Sect.~\ref{sec:flux_bands}. The intra-band spectral index ($\alpha_{LOFAR}$) 
compared to the LOFAR to NVSS spectral index ($\alpha_{LOFAR-NVSS}$) is shown in 
Fig.~\ref{fig:alpha}. The distribution of $\alpha_{LOFAR}$ has a median of 0.92 
with an uncertainty on this median of 0.27. This large uncertainty on the 
intrinsic value arises from the combined flux density scaling uncertainties in 
the three bands (see Sect.~\ref{sec:flux_bands}); flux density calibration 
changes would systematically shift all of the measured $\alpha_{LOFAR}$ in the 
field in the same direction. The value is in agreement with that found for the 
GLEAM survey \citep{HurleyWalker2017}. The standard deviation of the 
$\alpha_{LOFAR}$ distribution is 0.47.

\begin{figure}[htbp]
   \centering
\includegraphics[width=\linewidth]{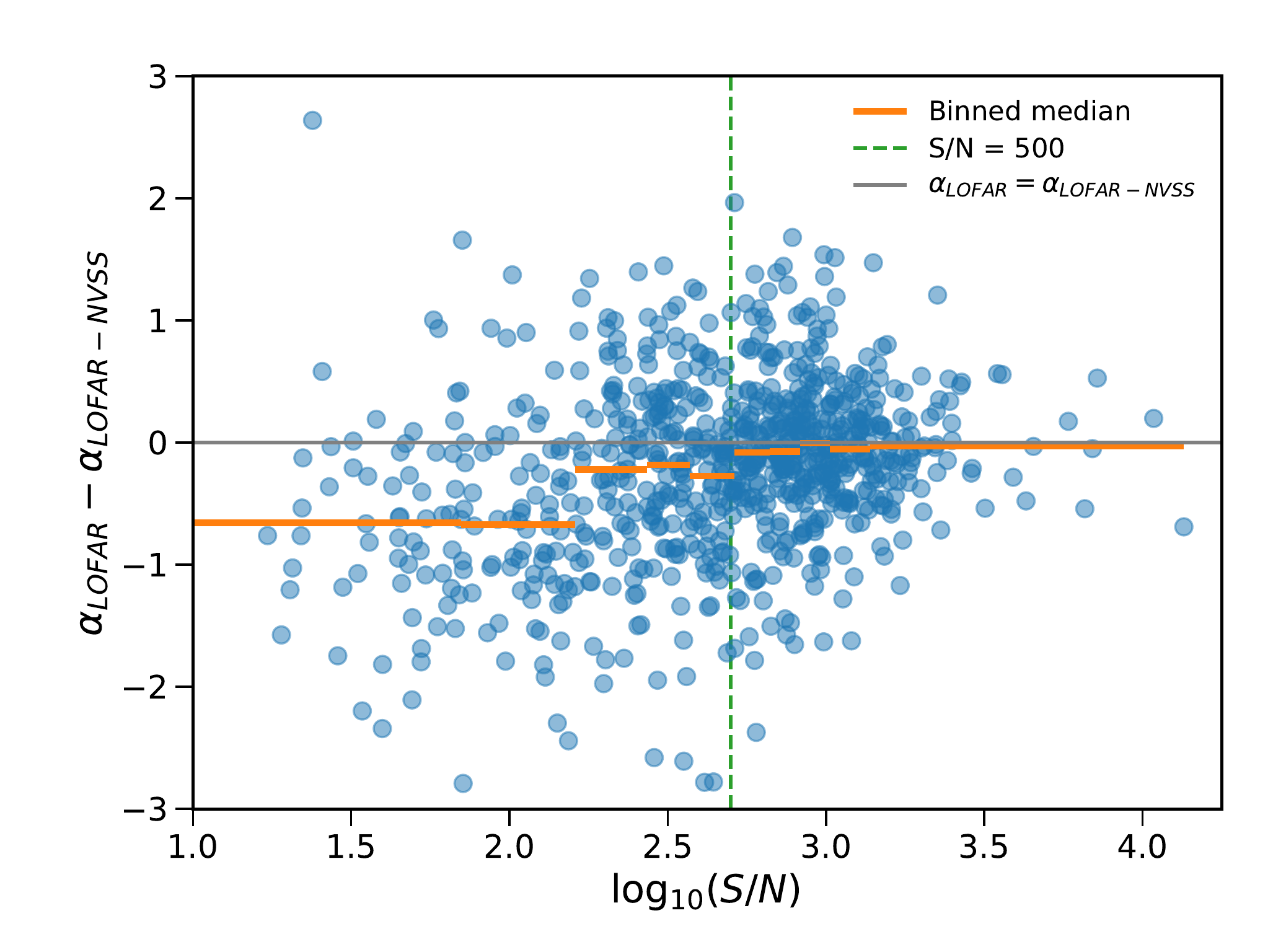}
   \caption{Difference between the intra-band spectral index ($\alpha_{LOFAR}$) 
and the NVSS to LOFAR spectral index ($\alpha_{LOFAR-NVSS}$) compared to the S/N 
of the cross-matched sources. The distribution of S/N is divided in deciles and 
the median of the difference is computed and shown as orange lines. The spectral 
indices are flatter (lower values of the difference) for low S/N likely due to 
incompletness effects. The distribution is close to zero (marked as a horizontal 
grey line) for sources with S/N$>500$ (shown as a vertical green dashed line).}
   \label{fig:alpha_sn}
\end{figure}

To obtain $\alpha_{LOFAR-NVSS}$, the NVSS sources cross-matched were required: 
a) to be at a distance of less than 3 degrees to the centre of the field; b) be 
cross-matched at a distance of less than 10 arcsecs; and c) have a major axis of 
less than 15 arcsecs. A final set of 799 sources were cross-matched to NVSS 
using these constraints. The distribution of $\alpha_{LOFAR-NVSS}$ (upper panel 
of Fig.~~\ref{fig:alpha}) has a median of 0.61 and a standard deviation of 0.47. 
The points in the main panel of Fig.~~\ref{fig:alpha} show strong scatter around 
the $\alpha_{LOFAR} = \alpha_{LOFAR-NVSS}$ line which can be related to the 
relatively high errors in their measurement. In Fig.~\ref{fig:alpha_sn} we check 
the relation between the S/N and the deviation from this line to test for 
incompleteness effects. The distribution of $\alpha_{LOFAR}-\alpha_{LOFAR-NVSS}$ 
below a S/N of 500 is skewed towards flatter spectral indices. This is likely to 
be produced primarily by the shallower depth of NVSS compared to LOFAR. 

A total least squares fit that takes into account possible correlated errors in 
the measurements \citep{Hogg2010} is applied to sources with S/N$>500$. We use 
the implementation of \textsc{astroML} \citep{astroML, astroMLText}. The fitting 
line is $\alpha_{LOFAR} = (0.78 \pm 0.03) (\alpha_{LOFAR-NVSS} - 0.73) - 0.09 
\pm 0.13$. The small shift downwards is related to the small difference in the 
median distribution of the $\alpha_{LOFAR}$ sources that are matched to NVSS 
with respect to the median of the total sample.

The intra-band spectral index is very sensitive to small variations in the flux 
density scale calibration. We studied the distribution of the spectral index 
with respect of the facet used in the calibration to check the accuracy of the 
flux density calibration in different facets. We used the 20 solution facets 
closer to the pointing centre as they cover the inner 3 degree radius where flux 
density calibration is more accurate as seen in Sect.~\ref{sec:variability}. We 
observe an scatter of about 0.2 in the distribution of the spectral indices that 
can be associated to the global facet to facet calibration. We checked that this 
level of scatter can be produced by a variation in the flux density scale of the 
order of a few per cent in the band images which is compatible with the accuracy 
of 6.5 per cent estimated in Sect.~\ref{sec:flux_scale}. Hence, apart from the 
possible global systematic shift of $\alpha_{LOFAR}$ which has a magnitude of 
0.27 there is an additional uncertainty within the field of the order of 0.2 
that must be taken into account. Nevertheless, despite this uncertainty, and the 
relatively large statistical uncertainties on the measured intra-band spectral 
index of individual sources, it can still be useful to detect sources with 
unusual spectral indices or general trends. For example, some sources in the 
lower right region of Fig.~\ref{fig:alpha} have a steep $\alpha_{LOFAR-NVSS}$ 
but inverted $\alpha_{LOFAR}$ indicating a likely spectral peak at a few hundred 
MHz \citep{Callingham2017}.

\subsection{Circularly polarized sources}
\label{sec:StokesV}

We detected two sources that present significant circularly polarized emission 
in the Stokes-V image. The first is the star CR Draconis which is an eruptive 
variable star, also detected by Callingham et al. (submitted) in the wider LoTSS 
survey.  It presents a relatively flat spectral index of $-0.8 \pm 0.7$ and the 
circularly polarized emission can be detected above the noise level in a couple 
of individual datasets, as will be explored in greater detail in Callingham et 
al. (in prep). The second source is the pulsar PSR J1552+5437. This millisecond 
pulsar was discovered using tied-array beam LOFAR observations 
\citep{Pleunis2017}. The pulsar is detected with an intra-band spectral index of 
$3.2 \pm 0.5$ which is similar to the value of \citet{Pleunis2017} within the 
error ($2.8 \pm 0.4$).


\section{Summary and conclusions}
\label{sec:conclusions}

In this paper we present the data and calibration procedure followed for the 
ELAIS-N1 field, the deepest of the LoTSS deep fields to date. The extended 
frequency coverage and configuration set-up arising from the co-observation with 
the EoR project of this field required a modified calibration process based on 
that used for the other deep fields \citepalias{Tasse2020}.

We applied to the field a careful flux density calibration method that took 
advantage of its good multi-frequency radio coverage. The uncertainty in the 
flux density derived from the flux density scale calibration method is at the 
level of 6.5 per cent. We provide the factors required to correct the flux 
density scale of the data.

We produced high resolution (6 arcsecs) Stokes-I images for the deep full 
dataset as well as for the individual datasets and spectral bands. Additionally 
low resolution (20 arcsecs) Stokes-I and V (uncleaned) images and QU datacubes 
were generated for the individual datasets. Source catalogues were produced for 
the high resolution Stokes-I images. Catalogues for the deep image, the spectral 
bands, and the individual datasets are provided. The final catalogue for the 
deep image contains 84862 sources.

We also examined in detail some properties of the data:
\begin{itemize}
 \item We analysed the reduction of the noise level as more data are added. The 
noise decreased almost as expected theoretically but there is some additional 
noise contribution, which is a combination of confusion noise, and additional 
noise that is likely to be associated with low-level residual calibration 
errors. The median of this extra contribution in the final deep image is at the 
16 per cent level.
 \item We studied the origin of the variation of flux density measurements 
between the individual datasets. The signal-to-noise ratio has the biggest 
impact as expected but we detect a remaining effect even at the higher 
signal-to-noise ratios. This effect has been linked to the quality of the data 
and calibration of the individual dataset. The primary beam correction seems to 
have an effect at large radii (larger than 3--4 degrees from the centre), but 
within 3 degree radius the flux density calibration is independent of radius. 
Variability also increases for more extended sources, particularly at low S/N, 
where the \textsc{PyBDSF} extracted-source parameters may vary from dataset to 
dataset. 
 \item We find an empirical fit to the contribution of the quality of the data 
and calibration of a dataset to the variability of the flux density 
measurements. Using this relation we find that an additional error of $3.3 \pm 
0.7$ per cent in the flux density measurement could be expected for the deep 
dataset. The final flux density uncertainty for ELAIS-N1, the Lockman Hole, and 
Bo{\"o}tes considering all the contributing factors is estimated to be $\sim$10 
per cent.
 \item We provide an analytic formula based on that of \citet{Franzen2015} and 
\citet{Shimwell2019} to estimate whether a source is likely to be extended or 
not. The estimation is dependent on the distance of the source to the centre of 
the field. Using this relation, we classify 9630 sources as extended at a 3 
sigma significance level.
 \item We provide values and an analysis of the intra-band spectral indices. A 
systematic global shift of the values is expected due to the uncertainty in the 
flux density scale calibration. This uncertainty follows a normal distribution 
with a sigma of 0.27. Apart from the error propagated from the uncertainty in 
the measurements, an additional error of the order of 0.2 in the intra-band 
spectral index is estimated.
\end{itemize}

This survey of ELAIS-N1 is the deepest radio survey of the region to date. We 
reach a sky density of sources of more than $\sim 5000$ sources per square 
degree in the central part (up to $\approx 2$ square degrees). The survey covers 
a wide area with a good spatial resolution. The number of sources detected in 
the central 2 square degrees is similar to that detected in total by VLA-COSMOS 
at 3~GHz in a similar sky area \citep[10973 versus 10830 in VLA-COSMOS at 3~GHz; 
see][]{Smolcic2017}. However, we pick up several times more sources over more 
than an order of magnitude larger sky area.

These radio data have been enriched with multiwavelength information and cross 
identifications \citepalias{Kondapally2020} as well as photometric redshifts 
\citepalias{Duncan2020}. This, combined with the quality of the radio data of 
ELAIS-N1, will enable a wide range of scientific studies of faint radio sources 
with high statistical significance and in representative environments. 

At the time of publication, more data have already been taken, and additional 
time is allocated for observation to reach a total exposure time of 500 hours by 
the middle of 2022. These data will allow us to approach the final depth of 
\mbox{$\sim$ 10 $\mu$Jy beam$^{-1}$}. 

\begin{acknowledgements}
The authors are indebted to Ger de Bruyn, sadly deceased, who played a critical 
role in the inception of the project.

We acknowledge the useful comments of the anonymous referee.

We acknowledge the joint SKA and AWS Astrocompute proposal call that was used to 
fund all the tests and calibration in the AWS infrastructure with the projects 
`Calibration of LOFAR ELAIS-N1 data in the Amazon cloud' and `Amazon Cloud 
Processing of LOFAR Tier-1 surveys: Opening up a new window on the Universe'. 
Cycle 0 and 2 data are available in the Registry of Open Data on AWS in: 
https://registry.opendata.aws/lofar-elais-n1/.

This paper is based on data obtained with the International LOFAR Telescope 
(ILT) under project codes LC0\_019, LC2\_024 and LC4\_008. LOFAR 
\citep{vanHaarlem2013} is the LOw Frequency ARray designed and constructed by 
ASTRON. It has observing, data processing, and data storage facilities in 
several countries, which are owned by various parties (each with their own 
funding sources) and are collectively operated by the ILT foundation under a 
joint scientific policy. The ILT resources have benefited from the following 
recent major funding sources: CNRS-INSU, Observatoire de Paris and Université 
d'Orléans, France; BMBF, MIWF-NRW, MPG, Germany; Science Foundation Ireland 
(SFI), Department of Business, Enterprise and Innovation (DBEI), Ireland; NWO, 
The Netherlands; The Science and Technology Facilities Council, UK.

This work was carried out in part on the Dutch national e-infrastructure with 
the support of SURF Cooperative through e-infra grants 160022 \& 160152.

JS, PNB, RKC, and RK are grateful for support from the UK Science and 
Technology Facilities Council (STFC) via grant ST/M001229/1 and ST/R000972/1.

MJH acknowledges support from the UK Science and Technology Facilities Council 
(ST/R000905/1)

VJ acknowledges support by the Croatian Science Foundation for the project 
IP-2018-01-2889 (LowFreqCRO).

HR acknowledges support from the ERC Advanced Investigator programme NewClusters 
321271.

MB acknowledges support from INAF under PRIN SKA/CTA FORECaST and from the 
Ministero degli Affari Esteri della Cooperazione Internazionale - Direzione 
Generale per la Promozione del Sistema Paese Progetto di Grande Rilevanza 
ZA18GR02.

MJJ acknowledges support from the UK Science and Technology Facilities Council 
[ST/N000919/1] and the Oxford Hintze Centre for Astrophysical Surveys which is 
funded through generous support from the Hintze Family Charitable Foundation.

RK acknowledges support from the Science and Technology Facilities Council 
(STFC) through an STFC studentship via grant ST/R504737/1.

IP acknowledges support from INAF under the SKA/CTA PRIN `FORECaST' and the 
PRIN MAIN STREAM `SAuROS' projects.

This research made use of \textsc{Astropy}, a community-developed core Python 
package for Astronomy (Astropy Collaboration, \citeyear{astropy}, 
\citeyear{astropy2}); \textsc{Ipython} \citep{IPython}; \textsc{matplotlib} 
\citep{matplotlib}; \textsc{numpy} \citep{numpy}; \textsc{pandas} 
\citep{pandas}; \textsc{scipy} \citep{scipy}, \textsc{TOPCAT} \citep{TOPCAT}, 
and \textsc{KERN} suite \citep{Kern}.

\end{acknowledgements}

\bibliographystyle{aa}
\bibliography{databasesol}

\begin{appendix}
\section{Sky model inception}
\label{sec:skymodel}
The first step of the processing consisted of the generation of an initial model 
for the ELAIS-N1 field. An initial model of the unresolved calibrator source 
87GB~160333.2+573543 was used for a direction-independent calibration of the 
flanking field that was centred in it. 

\begin{figure}[htbp]
   \centering
\includegraphics[width=\linewidth]{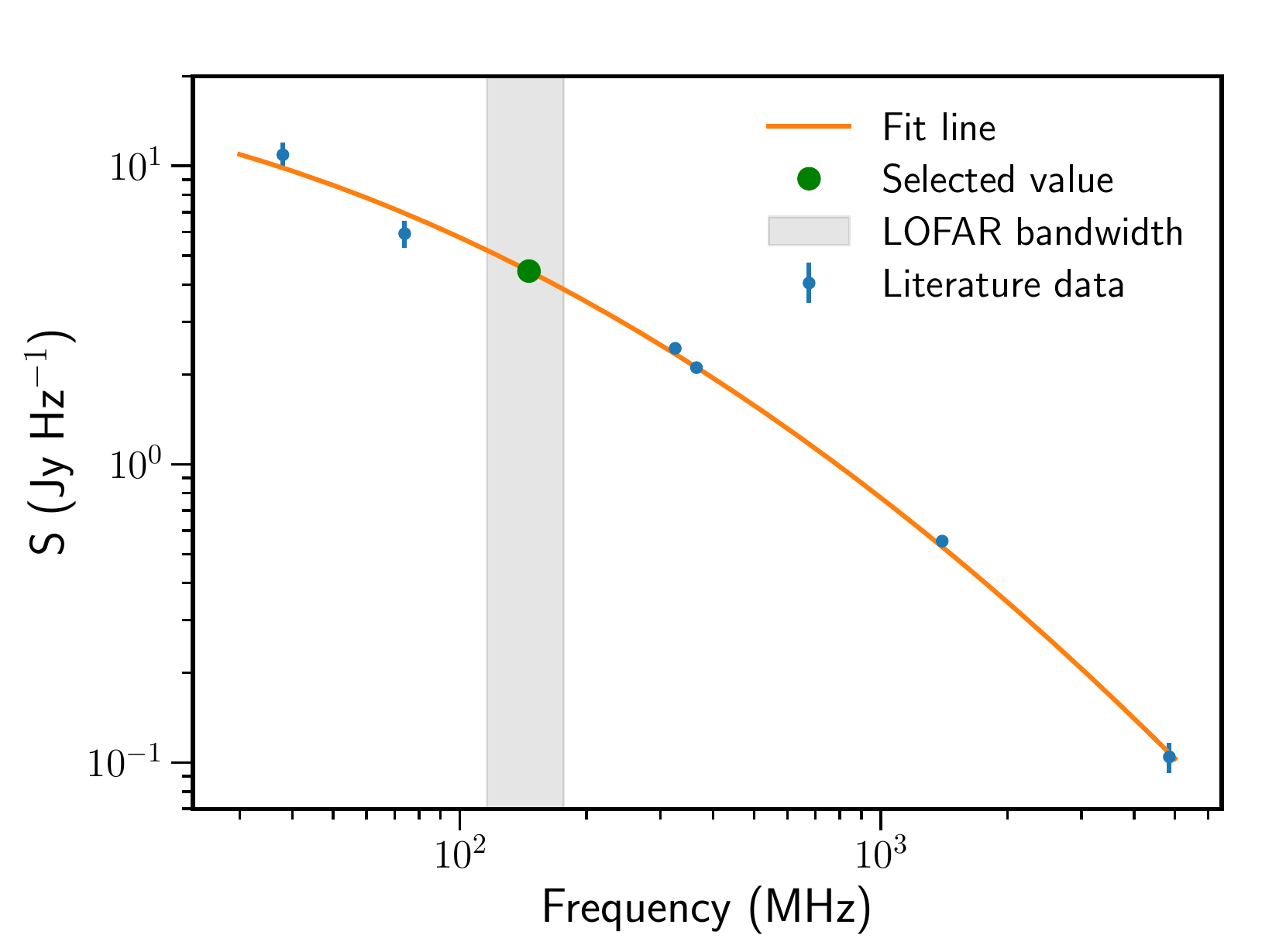}
   \caption{Determination of the flux density of the calibrator 
87GB~160333.2+573543 in 
the LOFAR band using 
literature data. The measurements from the literature and their respective 
errors are shown as blue dots. The vertical grey band represents the LOFAR 
bandwidth of the ELAIS-N1 data. The green dot marks the value of the flux 
density adopted at a frequency of 146~MHz.}
   \label{fig:87GBfit}
\end{figure}

We used data from several radio catalogues to characterize the flux, spectral 
index slope and curvature for the calibrator source. The data were taken from 
the 87GB, Texas Survey of Radio Sources, WENSS, VLSS and 8C, covering from 38 
MHz to 5~GHz. The measurements were fitted with a polynomial fit of second 
order, to determine that the flux density of the calibrator at 146 MHz is 
4.44~Jy with a spectral index of 0.72 and a curvature of $-0.22$. A sky model 
with a single source with these parameters was the one used as an initial input 
for the calibration of the flanking field. Fig.~\ref{fig:87GBfit} shows the 
result of the fit along the measurements and the location of the ELAIS-N1 LOFAR 
band. The error of the measurements was also considered in an ODR fit and the 
results were similar to the polynomial fit within a 0.2 per cent factor.

The gain solutions (amplitude and phase) for this flanking field were 
transferred to the ELAIS-N1 target field and a direction-independent calibration 
was applied to obtain a preliminary sky model of the ELAIS-N1 field. The mosaic 
observation with id. 000 was used because the frequency configuration for the 
flanking and target fields is the same and allowed a direct transfer of the 
solutions. The initial sky model was thereafter fed in as the input for the 
calibration of the dataset 003. A simple direction-independent calibration was 
performed, solving and correcting for the phase and amplitude gains. The 
calibration was executed in 8 different bands. A model was extracted from the 
images using \textsc{PyBDSF} \citep{Mohan2015}. The number of sources increased 
by a factor 10 due to the wider bandwidth used on the target field in this 
dataset.

\section{Infrastructure}
\label{sec:infrastructure}

The processing of the data required a combination of high throughput and high 
performance computing facilities. As the calibration pipeline was developed the 
computing and storage requirement changed mainly in two ways: a) the storage 
requirements diminished by applying optimized levels of averaging to the data 
and compression \citep{Offringa2016a}; b) the memory and scratch area size 
requirements increased as the third generation calibration software performance 
was tuned. The software installation also played an important role in the 
selection of the infrastructure as explained in \citet{Sabater2017}.

The computing cluster of the Instituto de Astrof\'{\i}sica de Andaluc\'{\i}a 
(Consejo Superior de Investigaciones Cient\'{\i}ficas) in Granada is composed of 
28 working nodes with 16 cores and 128 GB of memory. A shared filesystem between 
the nodes allowed the system to seamlessly access the data. The Cycle 0 datasets 
000 and 003 were processed in this cluster with the first direction-independent 
pipelines.

The grant of a SKA-AWS astrocompute project allowed the use of the Amazon Web 
Services cloud infrastructure. This mainly solved the problems with the 
installation and management of the software that was blocking the progress of 
the calibration at this stage. The run of the pre-factor pipelines for the 
cycles 0 and 2 of the data and several direction-dependent calibration tests 
were performed in this infrastructure \citep{Sabater2017}.

The \textsc{pre-factor} pipeline was adapted to work in the SURF-Sara GRID 
infrastructure for the calibration of the LoTSS data \citep{Mechev2017}. The use 
of a custom data model for the ELAIS-N1 field and the processing of the extended 
bandwidth was allowed with minimal modifications to the pipeline. Hence, the 
pre-processing of the Cycle 4 datasets was performed in this infrastructure.

Finally, the Cuillin cluster of the Institute of Astronomy of the University of 
Edinburgh was adapted to the requirements of the next generation 
direction-dependent pipelines. The cluster contains several nodes with 32 cores 
and 512 GB of memory. It also provides storage for the data which is accessible 
from the nodes and big enough scratch data areas to hold the intermediate data. 
The final direction-dependent calibration was performed in Cuillin.

\section{PyBDSF parameters}
\label{sec:pybdsfparameters}

The parameters used to 
extract the sources are similar to those used in LoTSS DR1 and are listed in 
Table~\ref{tab:pybdsf_params}. The primary beam uncorrected image, which has 
more uniform noise properties than the corrected one, was entered 
as the detection image (\texttt{detection\_image} parameter) to detect islands 
of emission whose characteristics were then derived from the primary beam 
corrected image. 

\begin{table}
\centering
\caption{PyBDSF parameters. The exact parameters entered to \textsc{PyBDSF} to 
extract the catalogues. Additionally, the parameter \texttt{detection\_image} 
was set to the path of the beam uncorrected image.}
\label{tab:pybdsf_params}
\begin{tabular}{ll}
\hline
 Parameter & value \\
\hline
\hline
rms\_box & \texttt{(150, 15)} \\
adaptive\_rms\_box & \texttt{True} \\
rms\_box\_bright & \texttt{(60, 15)} \\
thresh\_isl & \texttt{4.0} \\
thresh\_pix & \texttt{5.0} \\
adaptive\_thresh & \texttt{150} \\
atrous\_do & \texttt{True} \\
atrous\_jmax & \texttt{4} \\
flag\_maxsize\_fwhm & \texttt{0.5} \\
ini\_method & \texttt{'intensity'} \\
mean\_map & \texttt{'zero'} \\
group\_by\_isl & \texttt{False} \\
group\_tol & \texttt{10} \\
rms\_map & \texttt{True} \\
output\_all & \texttt{True} \\
output\_opts & \texttt{True} \\
flagging\_opts & \texttt{True} \\
\hline
\end{tabular}
\end{table}

\section{Flux scale corrections for the Lockman Hole and Bo{\"o}tes deep fields}
\label{sec:LHBoo}

\begin{figure*}[htbp] \centering 
\includegraphics[width=\linewidth]{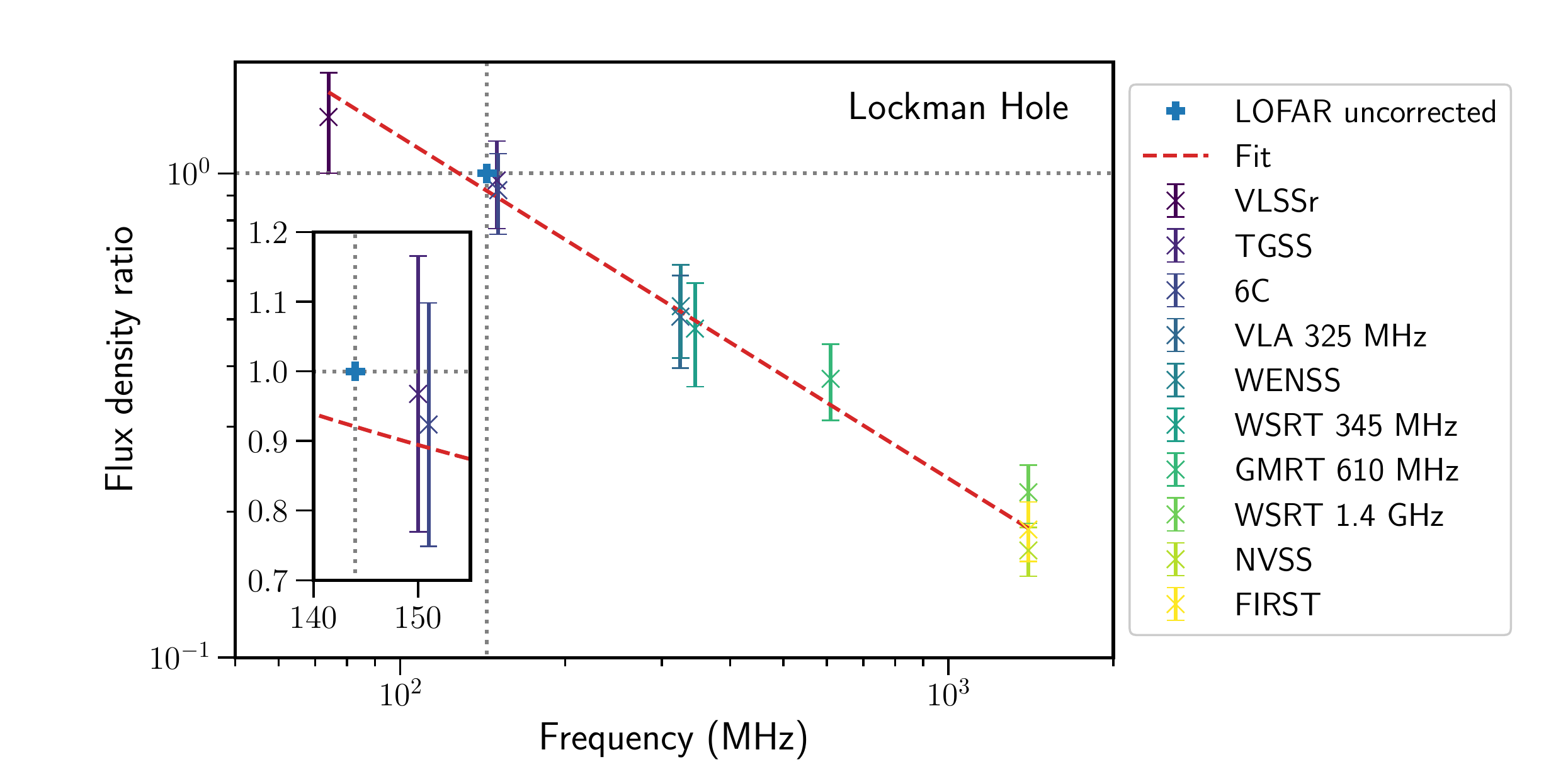} 
\includegraphics[width=\linewidth]{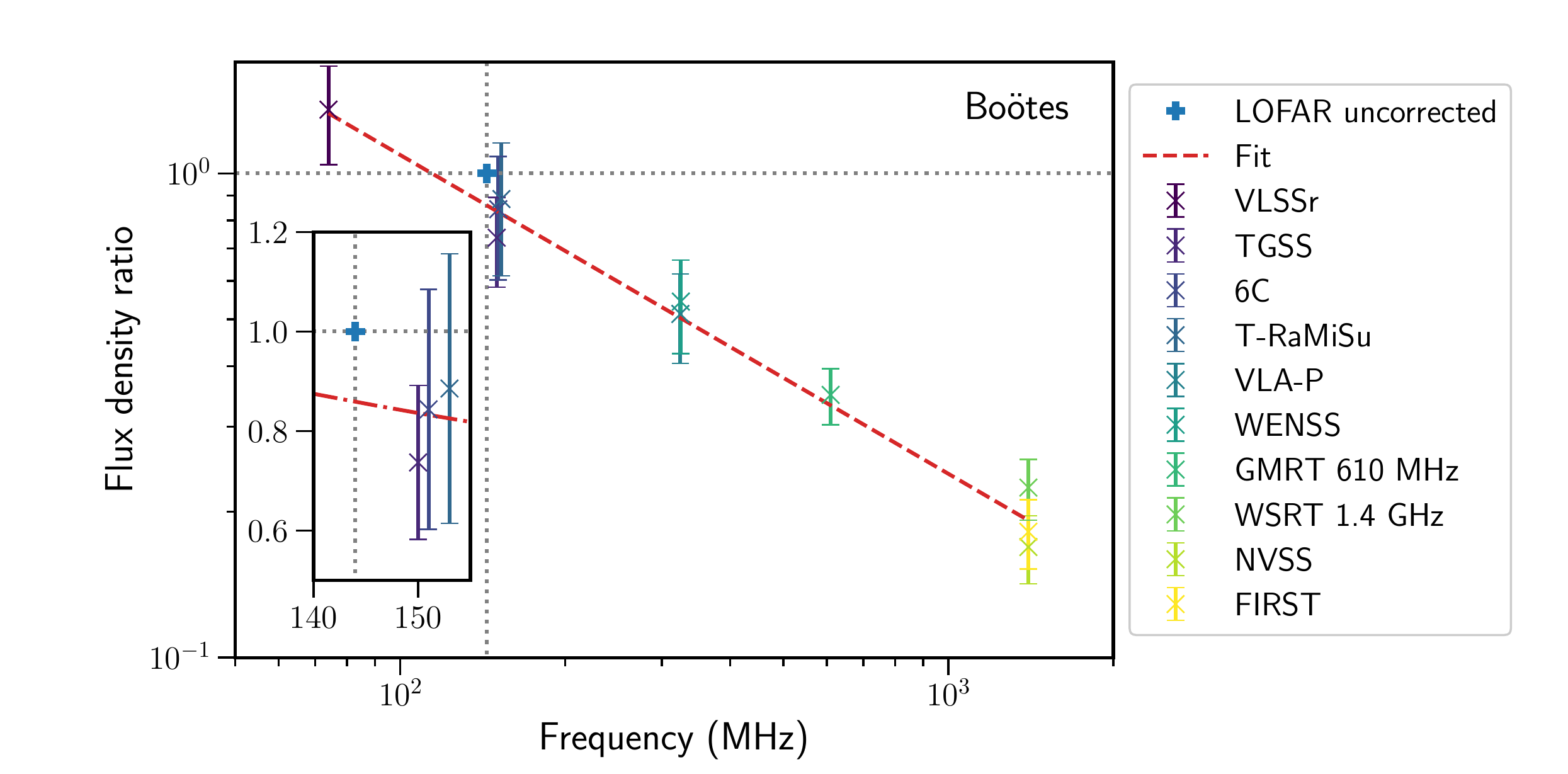} 
\caption{Calibration of the Lockman Hole (upper panel) and Bo{\"o}tes (lower 
panel) flux density scales. The default \textsc{DDF-pipeline} scale is set to 
unity and marked with a blue cross. The flux density ratios with respect to 
other surveys in the literature and their errors are shown in different colours 
and the fitting line is shown as a red dashed line. The inset shows a zoom view 
of the area close to the 144 MHz frequency of the observation.} 
\label{fig:fluxscaleLHBootes} \end{figure*}

We applied the method presented in Sect.~\ref{sec:flux_scale} to determine the 
flux density scale corrections to the Lockman Hole and Bo{\"o}tes LoTSS deep 
fields that are presented in \citetalias{Tasse2020}. The surveys used for the 
Lockman Hole are the VLSSr, TGSS, 6C, the VLA survey at 325~MHz by 
\citet{Owen2009}, WENSS, WSRT 345~MHz by \citet{Mahony2016}, the GMRT survey at 
610~MHz by \citet{Garn2008b}, the WSRT survey at 1.4~GHz by 
\citet{Prandoni2018}, NVSS, and FIRST. For the Bo{\"o}tes field the surveys used 
are VLSSr, TGSS, 6C, the T-RaMiSu survey at 153~MHz by \citet{Williams2013}, the 
VLA study at 325 MHz by \citet{Coppejans2015}, WENSS, the GMRT survey at 610~MHz 
by \citet{Coppejans2016}, the WSRT survey at 1.4~GHz by \citet{deVries2002}, 
NVSS, and FIRST. Fig.~\ref{fig:fluxscaleLHBootes} show the flux density ratios 
and the fit line. The linear fit gives values of $0.920^{+0.041}_{-0.039}$ for 
the Lockman Hole and $0.859^{+0.036}_{-0.034}$ for Bo{\"o}tes at a nominal 
frequency of 144~MHz. The linear fits were favoured over higher order polynomial 
fits by the AIC and BIC. The final uncertainty in the flux density scale 
calibration for the two fields is $~sim$10 per cent if the additional 
uncertainty term derived in Sect.~\ref{sec:variability} is included.

\end{appendix}

\label{lastpage}
\end{document}